\documentclass[sigconf,nonacm]{acmart}

\usepackage{textcomp}
\usepackage{graphicx}
\DeclareGraphicsExtensions{.png}
\usepackage{booktabs}
\usepackage{color}
\usepackage{colortbl}
\usepackage{multirow}
\usepackage{algorithm}
\usepackage{algpseudocode}
\usepackage{multirow}
\usepackage{pifont}
\usepackage{tikz}
\usepackage{xspace}
\usepackage{hhline}
\usepackage{subcaption}
\usepackage{soul}
\usepackage{enumitem}

\usepackage{makecell}
\usepackage{stfloats}

\usepackage{minibox}
\newcounter{observcntr}
\newcommand*{\observ}[1]{%
    \stepcounter{observcntr}%
    \begin{center}
    \vspace{2pt}
    \minibox[frame, rule=1pt,pad=3pt]{
        \begin{minipage}[t]{0.95\columnwidth}
        \textbf{Remark~\arabic{observcntr}:} \textit{#1}.
        \end{minipage}
    }
    \vspace{2pt}
    \end{center}
}
\newcommand{\subheading}[1]{\noindent{\textbf{#1}}}

\newcommand{\new}[1]{\textcolor{black}{#1}}

\definecolor{myhl}{rgb}{0.9,0.8,0}

\newcommand{\sysname}{\textsc{DeepTaster}\xspace}
\newcommand{\deepjudge}{\textsc{DeepJudge}\xspace}
\usepackage{filecontents}

\newcommand{\red}[1]{\textcolor{red}{}}

\AtBeginDocument{%
  }

\copyrightyear{2023}
\acmYear{2023}
\setcopyright{acmlicensed}\acmConference[ACSAC '23]{Annual Computer Security Applications Conference}{December 4--8, 2023}{Austin, TX, USA}
\acmBooktitle{Annual Computer Security Applications Conference (ACSAC '23), December 4--8, 2023, Austin, TX, USA}
\acmPrice{15.00}
\acmDOI{10.1145/3627106.3627204}
\acmISBN{979-8-4007-0886-2/23/12}

\begin{document}

\title[\sysname: Adversarial Perturbation-Based Fingerprinting to Identify Proprietary Dataset]{\sysname: Adversarial Perturbation-Based Fingerprinting to Identify Proprietary Dataset Use in Deep Neural Networks}


\author[Park et al.]{Seonhye Park$^{1}$, Alsharif Abuadbba$^{2}$, Shuo Wang$^{2}$, Kristen Moore$^{2}$, Yansong Gao$^{2}$,\\ Hyoungshick Kim$^{1}$, Surya Nepal$^{2}$,}
\def \authors{Seonhye Park, Alsharif Abuadbba, Shuo Wang, Kristen Moore, Yansong Gao, Hyoungshick Kim, Surya Nepal}
\affiliation{%
\institution{$^1$Sungkyunkwan University
\country{Republic of Korea}}
}
\email{{qkrtjsgp08, hyoung}@skku.edu}
\orcid{1234-5678-9012}
\affiliation{%
  \institution{$^2$CSIRO' Data61 
  \country{Australia}}
}
\email{{sharif.abuadbba, shuo.wang, kristen.moore, garrison.gao, surya.nepal}@data61.csiro.au}

%\authornote{.}

%-------------------------------------------------------------------
\begin{abstract}
%-------------------------------------------------------------------
%Training Deep Neural Networks (DNNs) requires extensive datasets and powerful computational resources. As a result, certain model owners restrict the redistribution of their DNNs without explicit permission. To protect DNN model ownership, watermarking schemes have been proposed, which embed confidential information within a DNN model and validate its presence to verify model ownership. However, these methods can compromise model utility and remain susceptible to watermark removal attacks. An alternative method, \deepjudge, was recently introduced, which measures the similarity between a suspect and a victim model without modifying the latter. While \deepjudge shows potential in overcoming watermarking limitations, it primarily targets scenarios where the suspect model's architecture mimics the victim's. 

Training deep neural networks (DNNs) requires large datasets and powerful computing resources, which has led some owners to restrict redistribution without permission. Watermarking techniques that embed confidential data into DNNs have been used to protect ownership, but these can degrade model performance and are vulnerable to watermark removal attacks. Recently, \deepjudge was introduced as an alternative approach to measuring the similarity between a suspect and a victim model. While \deepjudge shows promise in addressing the shortcomings of watermarking, it primarily addresses situations where the suspect model copies the victim's architecture. In this study, we introduce \sysname, a novel DNN fingerprinting technique, to address scenarios where a victim's data is unlawfully used to build a suspect model. \sysname can effectively identify such \new{DNN model theft attacks}, even when the suspect model's architecture deviates from the victim's. To accomplish this, \sysname generates adversarial images with perturbations, transforms them into the Fourier frequency domain, and uses these transformed images to identify the dataset used in a suspect model. The underlying premise is that adversarial images can capture the unique characteristics of DNNs built with a specific dataset. To demonstrate the effectiveness of \sysname, we evaluated the effectiveness of \sysname by assessing its detection accuracy on three datasets (CIFAR10, MNIST, and Tiny-ImageNet) across three model architectures (ResNet18, VGG16, and DenseNet161). We conducted experiments under various attack scenarios, including transfer learning, pruning, fine-tuning, and data augmentation. \new{Specifically, in the Multi-Architecture Attack scenario, \sysname was able to identify all the stolen cases across all datasets, while \deepjudge failed to detect any of the cases.}

 \end{abstract}




\maketitle

\section{Introduction}

%Deep neural networks (DNNs) have recently gained much attention from academia and industry because they have proved useful in numerous applications, including image recognition~\cite{wang2018cosface}, autonomous driving~\cite{luo2017traffic}, and medical image classification~\cite{zhang2019medical}. One of the reasons for their success and widespread utilization in various domains is that IT giants such as Google, IBM, Microsoft, and OpenAI have released their pre-trained DNN models to the scientific community to promote further research and scientific advancement. In many cases, pre-trained models have been built on huge datasets collected, processed, organized, and labeled by the organisation.

%Organisations that seek to monetise their proprietary DNN models can utilise cloud providers that offer Machine Learning as a Service (\texttt{MLaaS}). However, it is important to note that there are potential security risks associated with using \texttt{MLaaS}, such as the theft of DNN models or datasets~\cite{sun2021mind}. In particular, the dataset for \texttt{MLaaS} could be accessed and misused by a malicious insider. For instance, the recent data breach incident at ``Capital One'' highlighted the risks associated with insider attackers on the cloud, where an unauthorised insider could access users' data on the cloud server~\cite{Murphy19:cloud}. This incident demonstrates the possibility of dataset misuse from \texttt{MLaaS} providers, who could steal a proprietary dataset and use it for their own DNN models without the owner's permission.

Organisations looking to commercialise their proprietary Deep Neural Network (DNN) models via Machine Learning as a Service (\texttt{MLaaS}) must be wary of the potential security risks entailed, notably the potential theft of DNN models or datasets~\cite{sun2021mind}. DNN models are constructed on huge datasets, which are meticulously collected, processed, organised, and labelled, usually requiring significant expense. Therefore, ensuring the protection of model and dataset ownership becomes crucial. 

We need to consider several attack scenarios related to the model and data theft. A recent data breach at ``Capital One'' showed the risk of insider attacks on cloud platforms, where someone with unauthorised access could steal data stored on the cloud server~\cite{Murphy19:cloud}. This security incident highlights the potential misuse of datasets by insider attackers, who could covertly steal a proprietary dataset and incorporate it into their own DNN models without the owner's consent. Another risk is external attackers stealing a DNN model by querying it through \texttt{MLaaS} APIs. Recent studies (e.g., \cite{papernot2017practical,yu2020cloudleak,Truong2021data}) have shown that \new{DNN model theft attacks} can be effectively conducted, even within real-world services. %Consequently, it becomes crucial for DNN model owners to protect their models' intellectual property (IP) against such theft attacks.

%Many attack scenarios can be envisaged, such as model and data theft. For example, a data breach incident at ``Capital One'' highlighted the potential for data theft from insider attacks on cloud platforms, wherein an unauthorised insider accessed data stored on the cloud server~\cite{Murphy19:cloud}. Such a security breach emphasises the potential for dataset misuse by insider attackers who could stealthily steal a proprietary dataset and implement it within their own DNN models without acquiring the consent of the dataset owner.

%Another possibility is the theft of a DNN model by external attackers by querying the model via \texttt{MLaaS} APIs. Recent studies (e.g., \cite{papernot2017practical,yu2020cloudleak,Truong2021data}) have shown that DNN model stealing attacks can effectively be launched even in real-world services. Therefore, it would be necessary for the DNN model owners to protect the intellectual property (IP) of their own models from stealing attacks. 

Existing DNN intellectual property (IP) protection mechanisms fall into two categories: \emph{DNN watermarking} and \emph{DNN fingerprinting}. DNN watermarking involves embedding the owner's information (i.e., a watermark) into a proprietary model~\cite{uchida2017embedding,darvish2019deepsigns,adi2018turning,abuadbba2021deepisign,zhang2018protecting,chen2018deepmarks,chen2019blackmarks,jia2021entangled,szyller2021dawn}. Model ownership can be verified by extracting an identical or similar watermark from a suspected model. There have been many proposals for developing effective DNN watermarking schemes. However, DNN watermarking presents two inherent limitations: (a) DNN watermarking is invasive by design, as it necessitates modifications to the original DNN model to embed a watermark, potentially altering the model's behaviour~\cite{zhang2018protecting,wang2019neural}. (b) DNN watermarking lacks sufficient resilience to adversarial attacks~\cite{Xue2021DNN, Yan2022Cracking}. Aiken \textit{et al.}~\cite{Aiken2020Neural} showed that attackers could effectively manipulate neurons or channels in DNN layers that contribute to the embedded watermark for most state-of-the-art DNN watermarking schemes. Lukas \textit{et al.}~\cite{lukas2021sok} recently demonstrated that transfer learning could remove nearly all of the tested 11 watermarking schemes.

In contrast, DNN fingerprinting is \textit{non-invasive} by design as it leverages the unique characteristics (i.e., fingerprinting features) of each DNN model without modifying the model itself. A verifier can identify the model by examining these fingerprinting features~\cite{lukas2019deep,IPGuard}. Generally, a single fingerprinting feature is insufficient to identify a model constructed through model theft and adaptive attacks~\cite{chen2021copy}. Chen \textit{et al.}~\cite{chen2021copy} recently introduced \deepjudge, a state-of-the-art fingerprinting scheme that leverages multiple fingerprinting features to protect a model's copyright. However, as \deepjudge utilises fingerprinting features tied to the model's parameters, it may struggle to detect unauthorised use of the protected training dataset if a suspect DNN model comprises different parameters or uses a distinct model architecture. Moreover, \deepjudge is not sufficiently effective in detecting models constructed through \textit{transfer learning}~\cite{Torrey2010transfer}. Our experimental results indicate that \deepjudge's detection accuracy is significantly degraded for models constructed through transfer learning. Consequently, \deepjudge may fail to identify instances where a victim's data is illicitly used to construct a suspect model with an architecture that differs from the original model of the victim. To tackle such attack scenarios, we propose a novel DNN fingerprinting scheme dubbed \sysname. 

%The state-of-the-art DNN fingerprinting scheme, \deepjudge~\cite{chen2021copy}, is designed to detect the unauthorized use of a victim's DNN model where a suspect model's architecture is the same as a victim model's. 

% \begin{figure}[h!]
%     \centering
%     \includegraphics[width=.9\linewidth]{figure/intro.pdf}
%     \caption{New attack scenario in which a victim's dataset is stolen to build a suspect model.}
%     %\caption{Critical dataset intelligence stealing attack that is not considered by exisitng fingerprinting and watermarking approaches.}
%     \label{fig:motivation}
% \end{figure}

In this paper, we demonstrate that the spectra of gradient-based adversarial examples can be used to identify the characteristics of a dataset used to train a DNN model, particularly regarding the model's decision boundaries. The adversarial perturbation images generated by gradient-based attacks preserve both the dataset and model characteristics, indicating a commonality among models trained on the same dataset. Our empirical analysis reveals that adversarial images' characteristics are more distinctively conserved in the Discrete Fourier Transform (DFT) domain compared to the spatial domain. Adversarial images typically contain more noise than standard images, a consequence of the adversarial perturbation process. We have observed that these noises are more noticeable in the frequency and spatial domains. As depicted in Appendix~\ref{Appendix:domain}, adversarial examples derived from three different architectures (ResNet18, VGG16, and Densenet161) on identical datasets (CIFAR10, MNIST, or Tiny-ImageNet) display significant similarity within the DFT domain. In contrast, adversarial examples within the spatial domain appear completely blacked out, making them visually indistinguishable across datasets. Inspired by these findings, we introduce \sysname, a method for detecting DNN model theft attacks. \sysname generates multiple adversarial images with perturbations, transforms them into the DFT domain, and uses their statistical properties as features to identify the dataset used to train a suspect model. Our experimental findings confirm that \sysname can accurately identify the dataset used to construct a suspect model, given that the architecture of the suspect model is known in advance. Our key contributions are summarised as follows:\looseness=-1

%\sysname generates multiple adversarial images with perturbations, transforms them into the DFT domain, and utilises their statistical properties as features to identify the dataset employed within a suspect model. Our experimental findings confirm that \sysname can precisely identify the dataset used to construct a suspect model, even when the architecture of the suspect model deviates from that of the victim model. To our knowledge, \sysname is the first attempt to detect this new form of model stealing attack. Our key contributions are summarized as follows:

\begin{itemize}[leftmargin=*]
    \item \new{We introduce \sysname, a novel DNN fingerprinting method designed to identify known model architectures trained on stolen datasets.} \sysname generates adversarial images, transforms them into the DFT domain, and uses these transformed images to discern the unique characteristics of the dataset used to train a suspect model. 
    \item We evaluate the resilience of \sysname against \new{eight} attack scenarios, including multi-architectures, data augmentation, retraining, transfer learning, fine-tuning, pruning, transfer learning with data augmentation\new{, and transfer learning with pretrained model}. These evaluations are conducted across three datasets (CIFAR10, MNIST, and Tiny-ImageNet) and three model architectures (ResNet18, VGG16, and DenseNet161). Our experimental results indicate that \sysname considerably outperforms \deepjudge in most scenarios. For example, in the Multi-Architecture Attack scenario, \sysname successfully detected \new{all the stolen cases} across all datasets. In contrast, \new{\deepjudge failed to detect four attack cases including transfer learning and data augmentation for the CIFAR10 dataset}.
\end{itemize}

\section{Background}
%------------------------------------------------------------------
%This section provides background information on adversarial perturbations and the DFT, which are two key concepts in \sysname. 

%\subsection{Deep Neural Networks (DNNs)}

%A DNN classifier is a function $f: X \to Y$ that maps the input $x \in X$ to the probability $y \in Y$ that the input belongs to each class. DNN classifier consists of $L$ layers $\{l_1, l_2, ..., l_L\}$, each of which is a set of neurons $\{n_{L,1}, n_{L,2}, ..., n_{L,N_L}\}$. Here, the first layer $l_1$ is called the input layer, the last layer $l_L$ is called the output layer, and the rest $l_2,..., l_{L-1}$ are called the hidden layers. The parameters within hidden layers are called weights and biases. The neurons that compose each layer calculate the output by applying a linear function followed by a non-linear function called the activation function to the input sequentially. We then apply a softmax activation function $\sigma(\cdot)$ to output layer $f_L(\cdot)$ to convert likelihoods into probabilities for each predicted class. Training the above DNN classifier requires a loss function that can be optimised by gradient descent on all trainable weights and biases. An example of loss function is cross-entropy. A black-box deployment of a DNN classifier only exposes the API of the model. The user sends an input element $x \in X$, the server will query the model internally and respond with a confidence vector of $\sigma(f_L(x)) \in Y$.
%the full confidence vector $\sigma(f_L(x)) \in Y$.

\subsection{Adversarial Perturbation and Attack}

An adversarial perturbation refers to an intentionally created perturbation of an input sample that can lead to its misclassification by a machine learning model~\cite{FGSM,PGD,wang2022octopus}. Gradient-based adversarial attacks are well-known for generating such perturbations, including the fast gradient sign method (FGSM)~\cite{FGSM}. Gradient-based adversarial attacks generate a minimal perturbation to the input sample in a direction that most significantly impacts the prediction of the target classifier. This ``small modification,'' which might be as subtle as changing a single pixel's color, can potentially disrupt the model's decision boundaries. For \sysname, we have chosen to use FGSM for its simplicity and satisfactory performance in generating adversarial images. We employ Foolbox~\cite{foolbox}, a commonly used library, to facilitate our experiments.

FGSM is a gradient-based adversarial attack algorithm~\cite{FGSM}. Let us consider $x$ as the original image and $\bigtriangledown$ as a slight perturbation applied to $x$, which leads to the creation of an adversarial sample, denoted as $\bar{x}$. The training process seeks to maximize the loss function $J(x,y)$ to derive the adversarial sample $\bar{x}$, which no longer belongs to class $y$. The entire optimization process has to fulfill the $L_{\infty}$ constraint $\left \| \bar{x}-x \right \|_{\infty} \leq \epsilon$. Accordingly, FGSM adversarial examples can be produced using the following equation:

\begin{equation}
\label{eq.fgsm}
    \bar{x} = x+ \epsilon \cdot sgn\left ( \bigtriangledown_x J (f(x),y)\right )
\end{equation}

Here, $sgn$ is the sign function, $J(f(x),y)$ is the loss function of the model's prediction for input $x$ and the true label $y$, and $\epsilon$ is a small constant which controls the magnitude of the perturbation.

\subsection{Discrete Fourier Transform (DFT)}

The Discrete Fourier Transform (DFT) is a tool used to transform a sequence of numbers $\{x_0, x_1, ..., x_N\}$ in the time domain into another sequence of numbers $\{y_0, y_1, ..., y_N\}$ in the frequency domain. The transformation is achieved through the equation $y_k=\sum_{n=0}^N x_n \cdot e^{-\frac{i2\pi}{N}kn}$. Upon application to an image, the DFT translates its spatial content into a frequency spectrum. This spectrum can be represented as an image, often revealing patterns imperceptible in the spatial domain. 

Adversarial images are typically noisier than natural images. These noises are a byproduct of the adversarial perturbation process. We found that the frequency domain shows these noises more clearly than the spatial domain. This is aligned with the findings of Harder \textit{et al.}~\cite{harder2021spectraldefense}, who showed that small changes in the spatial domain are hard to detect because they appear random. However, the same changes can lead to systematic changes in the frequency domain, which are then detectable. Our experimental results show that observing adversarial perturbations in the Fourier domain can be more advantageous than the spatial domain in identifying DNN models built with a specific dataset.

\section{Threat Model}\label{sec:threat_model}
%-------------------------------------------------------------------------------
%\sharif{Still fixing the threat model section.}
% \begin{figure}[t!]
%     \centering
%     \includegraphics[width=\linewidth]{figure/attack3.pdf}
%     \caption{\centering{Possible Attack Scenarios.}}
%     \label{fig:attack}
% \end{figure}

%\seonhye{How about changign the order of section3 and section4? I think it'd be better to explain Deeptaster, explain attack scenarios, move on to experimental results} \sharif{Usually security ppl will focus on threat models first before reading the models, so I would suggest we keep that flow:)}
%In this section, we describe the threat model that we consider in our study.

\subheading{Overview.} We examine the issue of dataset leakage in the context of DNN models. High-profile incidents, such as the theft of over 5.6 million fingerprint records from the US Office of Personnel Management (OPM)~\cite{gootman2016opm} and the ``Capital One'' data breach~\cite{Murphy19:cloud}, have highlighted the risks posed by both external and insider threats. If a dataset is leaked, an adversary could use it to create a new DNN model or improve an existing one. In another scenario, an adversary could steal models from the victim's private cloud or use the MLaaS API of the victim model to extract them. In this case, the adversary could use fine-tuning, pruning, and transfer learning to improve the performance of the stolen model and hide the signs of theft. Therefore, we considered the following threat models.

\subheading{Assumptions.} We make the following assumptions: (a) The adversary can steal the dataset used to train a victim's DNN model or the model itself. Malicious insiders can access unencrypted databases, while external threats could employ advanced techniques (e.g., SQL injection) to obtain plaintext data from databases. (b) The adversary aims to build a model with the illegally obtained dataset and evade copyright verification. (c) The surrogate model, as crafted by the adversary, provides sufficient accuracy that the adversary can profit from its sale or commercialisation.

\subheading{Settings.} In our experiments, we considered the following different adversarial settings. Table~\ref{tb:adveserialsettings} summarises these attacks along with the assumptions about the attacker's access level. None of the existing DNN IP protection mechanisms has considered attack scenarios (1), (2), (7)\new{, and (8)}.

\begin{table}[h]
\caption{Summary of Adversarial Settings. The ``Access'' column indicates the access required by an attacker.}\centering
\resizebox{3.2in}{!}{
\begin{tabular}{llcc}
\hline
\multirow{2}{*}{N} & \multicolumn{1}{c}{\multirow{2}{*}{{\textbf{Attack}}}} & \multicolumn{2}{c}{{ \textbf{Access}}}  \\ \cline{3-4} 
                    & \multicolumn{1}{c}{}                                       & \multicolumn{1}{c}{Dataset}      & Model   \\ \hline
1                   & Multi-Architecture Attack (MAA)                             & \multicolumn{1}{c}{Full}         & None \\ \hline
2                   & Data Augmentation Attack (DAA)                              & \multicolumn{1}{c}{Full%Full/Partial
} & None \\ \hline

3                   & Model Retraining Attack (SAA)                               & \multicolumn{1}{c}{Partial}      & None    \\ \hline
4                   & Transfer Learning Attack (TLA)                              & \multicolumn{1}{c}{None}      & Full    \\ \hline
5                   & Model Fine-tuning Attack (MFA)                              & \multicolumn{1}{c}{Partial}      & Full    \\ \hline
6                   & Model Pruning Attack (MPA)                                  & \multicolumn{1}{c}{Full}      & Full    \\ \hline
7                   & Data Augmentation and Transfer Learning Attack (DATLA)                            & \multicolumn{1}{c}{Full%Full/Partial
} & Full \\ \hline
8                   & \new{Transfer Learning with Pretrained mode Attack (TLPA)}                            & \multicolumn{1}{c}{\new{Full}%Full/Partial
} & \new{None}\\ \hline
\end{tabular}}
\label{tb:adveserialsettings}
\end{table}

%Appendix~\ref{Appendix:Attack} summarises these attacks, along with the assumptions about the attacker's access level. 

\subheading{(1) Multi-Architecture Attack (MAA).} 
The attacker steals the victim's dataset. The attacker trains a model with an architecture different from the original victim model using the stolen data.

\subheading{(2) Data Augmentation Attack (DAA).} The attacker steals the victim's dataset and creates a new one by combining the stolen data with new data. The attacker is aware of the structure of the victim's model. The attacker trains a model that mirrors the victim's model structure using the combined data.

\subheading{(3) Same Architecture Attack (SAA)~\cite{lukas2021sok}.} The attacker steals part of the victim's dataset. The attacker is aware of the structure of the victim's model. The attacker trains a model mirroring the victim's model structure using the stolen data.

\subheading{(4) Transfer Learning Attack (TLA)~\cite{lukas2021sok}.} The attacker steals the victim's model. The attacker uses transfer learning to fine-tune it on another dataset.

\subheading{(5) Model Fine-tuning Attack (MFA)~\cite{lukas2021sok}.} The attacker steals part of the victim's dataset and the victim's model. The attacker fine-tunes the model using the stolen dataset. 

\subheading{(6) Model Pruning Attack (MPA)~\cite{lukas2021sok}.} The attacker steals the victim's model. The attacker prunes the model and redistributes it.

\subheading{(7) Data Augmentation and Transfer Learning Attack (DATLA).} The attacker steals the victim's dataset and model. They then create a new dataset by combining the stolen data with new data. The attacker uses transfer learning to fine-tune the stolen model using the combined dataset.

\subheading{\new{(8) Transfer Learning with Pretrained model Attack (TLPA).}} \new{The attacker steals the victim's dataset. The attacker uses transfer learning to fine-tune the pretrained model with another dataset on the victim's dataset.}

\section{\sysname System Design}\label{sec:system_design}
%\seonhye{put full name of attack in the figure}
%-------------------------------------------------------------------------------
In this section, we present \sysname, a DNN IP tracking tool that verifies whether an attacker's model has been trained using a victim's dataset or model. We first present an overview of the system design and then detail the system's three main components: adversarial perturbation generation and transformation, classifier generation, and verification.

%\shuo{Move the requirements to overview?}
%\subsection{Design Requirements} 

%Protecting dataset IP presents new challenges compared to protecting model-dependent IP, which existing watermark and fingerprinting schemes can potentially address. To address these challenges, we establish the following design criteria for dataset IP protection mechanisms:

%\begin{enumerate}[leftmargin=*] 
%\item \textbf{Robustness:} The protection should resiliently identify the dataset's characteristics when the model architecture changes.
%\item \textbf{Fidelity:} The ownership protection and verification procedure should not detrimentally impact the model utility. 
%\item \textbf{Efficacy:} The verification process should be performed accurately. 
%\item \textbf{Efficiency:} The verification process should be performed efficiently. 
%\end{enumerate}
%-------------------------------------------------------------------------------
\subsection{\sysname Overview}
% \shuo{flow of these three components according to the criteria above}

\begin{figure}[ht!]
    \centering
    \includegraphics[width=.92\linewidth, trim=0cm 0.2cm 0cm 0cm, clip=true]{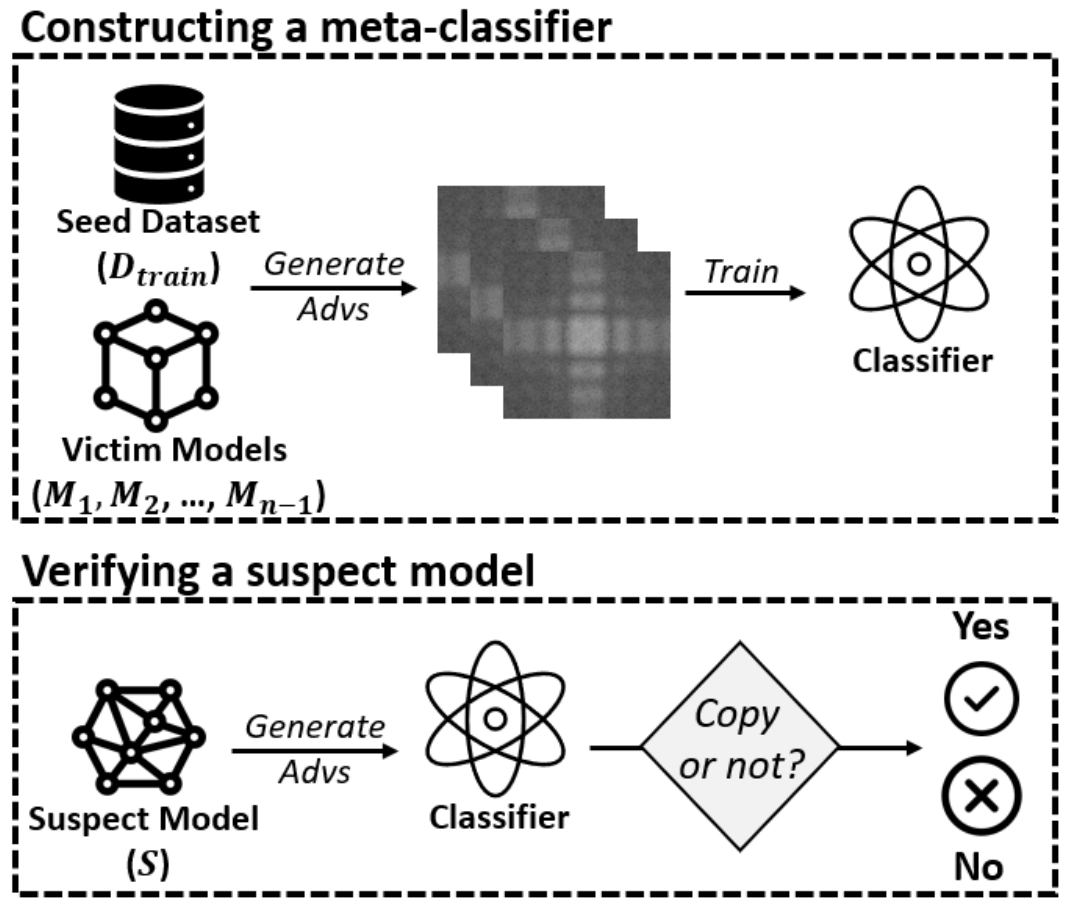}
    \caption{Overview of \sysname.}
    \label{fig:overview1}
\end{figure}

\sysname operates in two stages: (a) constructing a classifier using adversarial images in the DFT domain from a set of representative models trained on the victim dataset, and (b) verifying a suspect model by testing it using its adversarial images in the DFT domain. Figure~\ref{fig:overview1} provides a schematic representation of this process.\looseness=-1

%Further details of each step are explained in the following subsections. The generation of adversarial images in the DFT domain is a critical component of \sysname. Therefore, we will begin by describing this process first.

%As depicted in Figure~\ref{fig:overview1}, \sysname consists of the following 2-step process: %(a) the generation of adversarial perturbation samples and their translation to the Fourier frequency domain using DFT, 
%(a) the creation of a meta-classifier that is trained on the spectra (i.e., DFT samples) in order to distinguish the dataset intelligence, and (b) verification of the suspect model by generating adversarial perturbation samples from it and then testing them using the meta-classifier. The generation of adversarial perturbation samples and their translation to the Fourier frequency domain using DFT is 

%Details of each step are described in the following subsections.

%The flow of the proposed \sysname is summarized as follows:
%Adversarial perturbations obtained from different commonly-adopted models are used as carriers of the dataset intelligence. The meta-classifier identifies common characteristics between the carriers as a fingerprint of the dataset intelligence. The adversarial perturbation of the suspect model is used in conjunction with the well-trained meta-classifier for verification.

%------------------------------------------------------------------------------

\subsubsection{Adversarial Perturbation Generation and Transformation}\label{sec:adversarial}

%\shuo{1. FGSM loss function;}

Given a victim dataset, the dataset proprietor uses Algorithm~\ref{alg:adveserial} to generate adversarial images from a target model $M$ constructed with the dataset. That is, the FGSM attack is performed on the target model $M$ to generate adversarial images as dictated by Equation~\ref{eq.fgsm}.

\begin{figure}[!h]
    \centering 
    \begin{tabular}{cc}
        \includegraphics[width=.86\linewidth]{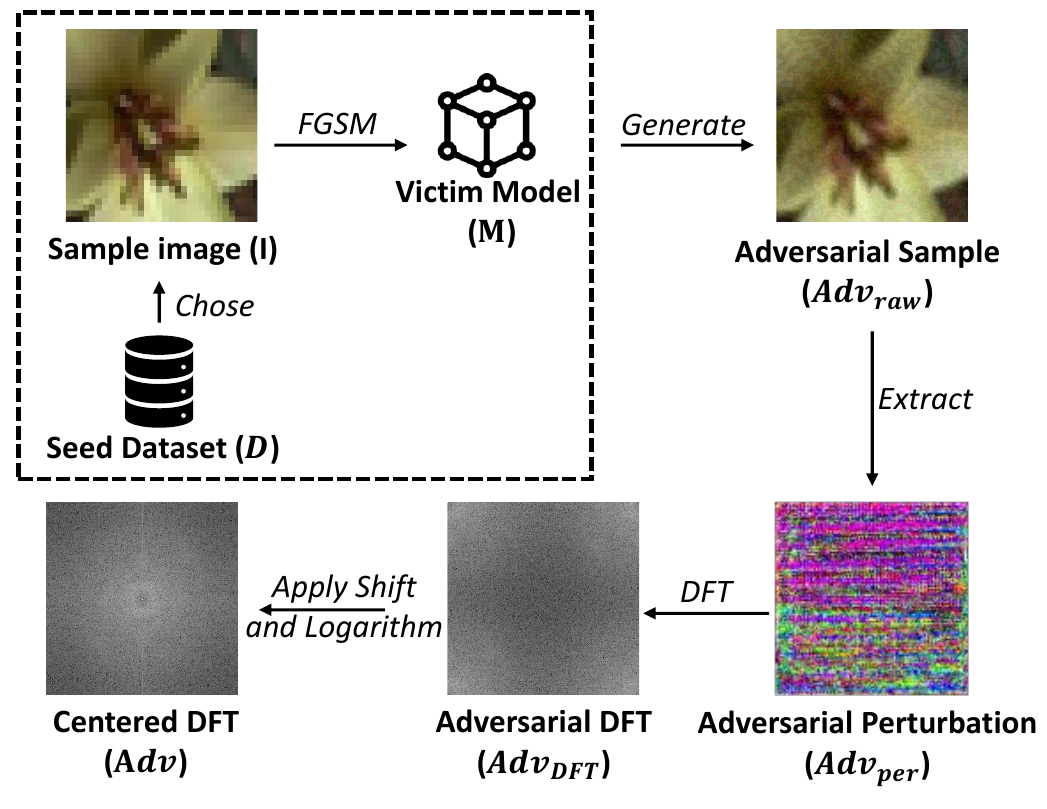}&
    \end{tabular}
    \caption{Adversarial Image Generation.}
    \label{fig:advGenTransform}
    \centering
\end{figure}

\begin{algorithm}[!h]
\caption{Adversarial Image Generation and Transformation.}\label{alg:adveserial}
\begin{flushleft} 
\textbf{Input}: Sample image $I$ and target model $M$ \\
\textbf{Output}: Adversarial DFT image $Adv$ 
\end{flushleft} 
\begin{algorithmic}[1]
\Procedure{$GenerateAdv$}{$M$,$I$}
\State $Adv_{raw} \gets FGSM(M,I)$
\State $Adv_{per} \gets Adv_{raw} - I$
\State $Adv_{DFT} \gets FourierTransform(Adv_{per})$
\State $Adv \gets Log(Shift(Adv_{DFT}))$
\State return $Adv$
\EndProcedure
\end{algorithmic}
\end{algorithm}
% \begin{equation}
%     Adv_{raw}=I+\epsilon \cdot sign(\Delta_IJ(\theta,I,y))
% \end{equation}
% \seonhye{\hl{Shuo, please check my comments!}}
% Here, $\theta$ is the model parameters and $J$ is the loss function.
%We use the Foolbox tool\cite{foolbox} with the specific epsilon value of $0.03$ and $l^2$-norm to conduct the attack.
We use the Foolbox library~\cite{foolbox} to generate adversarial images using the FGSM method. We use the $\ell_2$-norm metric to calculate the distance and set the epsilon value to $0.09$. The same seed images are used for all victim models tested to generate adversarial images. When the seed image domain differs from the victim model's image domain, we apply the FGSM method by re-labeling the seed images with the prediction value of the model. Adversarial images are selected when they successfully fool a victim model into producing an incorrect prediction. The adversarial perturbation $Adv_{per}$ is the pixel-wise difference between the original image $I$ and its adversarial image $Adv_{raw}$. We then transform the adversarial image $Adv_{raw}$ into the frequency domain, resulting in the adversarial DFT image $Adv_{DFT}$. To more accurately capture the characteristics of the dataset intelligence using the adversarial DFT images, we generate a \emph{centered} DFT image, denoted as $Adv$, by applying a shift operation and a logarithm function sequentially to the adversarial DFT image $Adv_{DFT}$. Figure~\ref{fig:advGenTransform} 
illustrates the process of generating a centered DFT image. However, for improved readability, we will refer to $Adv$ as the adversarial DFT image throughout the rest of the paper instead of using the term centered DFT images.
%\sharif{This seems cut off here. Please finish this part. You may need to be specific about the FGSM loss function and perturbation parameters selected! Just put details and we can help you rewrite that.. do not feel scared :)!}

%\sharif{We may need to draw Figure~\ref{fig:advGenTransform}, then rewrite the below text aligned with the steps in the figure -- if they are correct :). Fix them if not.}

%-------------------------------------------------------------------------------
\subsubsection{Classifier Construction}\label{sec:meta_classifier}
We have constructed a one-class classifier that is trained on adversarial DFT images generated from multiple model architectures, all of which are trained on the same dataset. Our goal is to build a robust detector that can determine whether a stolen dataset has been used to build a target model, even when the target model differs from the original one.
%To ensure robust dataset intelligence characteristics are captured, we develop a one-class meta-classifier that is trained on adversarial DFT images generated from multiple model architectures, each of which is trained on the victim dataset. The intuition here is to \textit{build a resilient detector that can efficiently recognise the stolen dataset intelligence, even when the adversary changes the model architecture or transfers the intelligence to other model as an adaptive attack strategy.}   
%To reduce the effort required to validate the model, victim can generate meta-classifiers only using the models that the victim has. To do this, we use one class classification (OCC) as Meta-classifier. One class classifier is trained on only target data and give output of the similarity value between input data and train data. 
To achieve this goal, we have chosen the Deep Support Vector Data Description (DeepSVDD) model~\cite{Ruff2018Deep} as our classifier from among several one-class classification models. DeepSVDD is an anomaly detection model that is widely used in different domains. It uses a deep neural network structure to find a hypersphere in the feature space. Therefore, we use DeepSVDD to extract data features from the adversarial discrete Fourier transform (DFT) images for models trained on a specific (victim) dataset. These features can then be used effectively by DeepSVDD to identify such models.

We construct a one-class classifier for \sysname using a seed dataset $D$ and a set of victim models $V$, which are trained on the victim dataset. To determine the optimal threshold of the classifier, we use a set of benign models $B_{val}$ trained on a different dataset than $V$. We split $D$ into two parts: $D_{train}$ and $D_{val}$. Similarly, we split $V$ into $V_{train}$ and $V_{val}$. We use $D_{train}$ and $V_{train}$ to train the one-class classifier and $D_{val}$, $V_{val}$, and $B_{val}$ to optimize the classifier's threshold $\tau$.\looseness=-1

To train the classifier, we generate adversarial samples $Adv_{train}$ from $D_{train}$ using Algorithm~\ref{alg:adveserial} for the victim models in $V_{train}$. We then use $Adv_{train}$ to train a DeepSVDD model as a one-class classifier.

To validate the classifier, we generate adversarial samples $Adv_{val}$ from $D_{val}$ for the victim models in $V_{val}$ using Algorithm~\ref{alg:adveserial}. We then evaluate the classifier's ability to distinguish between $V_{val}$ and $B_{val}$ using $Adv_{val}$ to determine the threshold $\tau$. The classifier is used to check whether a suspect model is built on $V$ or another dataset, and we set the threshold $\tau$ to maximize the classifier's accuracy.

For testing, the one-class classifier uses the threshold $\tau$ to verify if a suspect model is built on the victim's dataset. The details of the model verification process are given in Section \ref{sec:verification}. Using a larger set of victim models with different architectures leads to higher accuracy for the one-class classifier.

We surmise that any random images could be used as seed images for the victim model. In our experiments, we randomly selected samples from CIFAR10 as seed images to generate adversarial images for models built on not only CIFAR10 but also other datasets such as MNIST and Tiny-ImageNet. However, analyzing the effects of seed samples and developing an algorithm to select effective seed samples would be an interesting area for future work.

\subsubsection{Verification}\label{sec:verification}

We test whether a suspect model $S$ was built using the victim dataset with the constructed one-class classifier using $D_{val}$. This dataset $D_{val}$ is used both for optimizing the threshold and generating the adversarial DFT images ${Adv}_{test}$ for verification. Algorithm~\ref{alg:val} provides a detailed description of the verification procedure of \sysname. The algorithm inputs include the classifier $\mathcal{M}$, its threshold value $\tau$, the dataset $D_{val}$, and a suspect model $S$. The output is a verification result indicating whether the suspect model contains part of the victim dataset.

\begin{algorithm}[!ht]
\caption{Validation using \sysname.}\label{alg:val}
\begin{flushleft} 
\textbf{Input}: Classifier $\mathcal{M}$, the threshold $\tau$, the dataset $D_{val}$, and the suspect model $S$.\\   
\textbf{Output}: Verification results
\end{flushleft} 

\begin{algorithmic}[1]
\State $Adv_{test} \gets GenerateAdv({S, D_{val}})$

\State $X \gets 0$
\State $k \gets len(Adv_{test})$
\While{$k \neq 0$}
\State $X \gets X +\mathcal{M}(Adv_{test}[k])<=\tau$
\State $k \gets k - 1$
\EndWhile
\If{$X > len(Adv_{test}) * \frac{1}{2}$}
    \State $S$ is a stolen model
\Else
    \State $S$ is a benign model
\EndIf
\end{algorithmic}
\end{algorithm}

In order to verify the suspect model $S$, we initially generate the adversarial DFT images from $D_{val}$ for $S$ using the steps described in Section~\ref{sec:adversarial}, as shown in lines 1--3. Next, we put these images to the test by feeding them into the one-class classifier, as shown in lines 4--7. If the fraction of samples that fall below the classifier's threshold $\tau$ exceeds 50\%, we classify the suspect model as having been trained on stolen data, as shown in lines 8--12. The default criterion for \sysname is set at 50\% to simplify the decision-making process and ensure an unbiased assessment. The ``\new{theft image detection rate}'' refers to the proportion of tested adversarial samples that fall below the classifier's threshold $\tau$.

%\sharif{You might say why .. I recall that we talked about the threshold calculation should not be on the evaluation dataset or something like that.} \seonhye{I rewrite them, Please check again!}

%-------------------------------------------------------------------------------
\section{Experiments}
\label{sec:Experiments} 
%-------------------------------------------------------------------------------
We implemented \sysname as a self-contained toolkit in Python. In this section, we evaluate the performance of \sysname against an extensive list of \new{eight} attacks mentioned in Section~\ref{sec:threat_model}. Some of these attacks, such as fine-tuning and pruning, are well-studied in watermarking~\cite{lukas2021sok}. We also examine \sysname against more challenging adaptive attack scenarios such as transfer learning, retraining, and the most challenging -- multi-architecture -- which has never been considered before in the literature. To ensure the generalisability of \sysname, we generate three classifiers that track CIFAR10, MNIST, and Tiny-ImageNet. We also compare our results to the best state-of-the-art fingerprinting technique, \deepjudge~\cite{chen2021copy}.

\subsection{Experimental Setup}\label{sec:experiment_setup}
\subheading{Datasets and victim models.} We use four datasets, including CIFAR10~\cite{cifar10}, MNIST~\cite{Lecun1998Gradient}, Tiny-ImageNet~\cite{le2015tiny}, and ImageNet~\cite{ImageNet}. The first three datasets are used as both victim and suspect datasets. The ImageNet dataset is used as a suspect dataset only. All datasets are image classification datasets with a varying number of classes, ranging from 10 classes in CIFAR10 and MNIST to up to 1000 in ImageNet, as described in Appendix~\ref{App:experiment_setup}. We note that we use only half of the Tiny-ImageNet dataset (i.e., 100 classes) to run the experiments in order to reduce the experimental computation time.

%\begin{table}[h!]
% \centering
% \caption{Description of the datasets for experiments.}
% \begin{tabular}{lll}
% \hline
% Dataset & \# Classes & Usage \\ \hline
% CIFAR10 & 10& Victim / Suspect\\ %\hline
% MNIST & 10 & Victim / Suspect\\ %\hline
% Tiny-ImageNet & 100 & Victim / Suspect\\ %\hline
% ImageNet& 1000& Suspect \\ \hline
% \end{tabular}
% \label{tb:dataset}
% \end{table}
% \begin{table}[h!]
% \centering
% \caption{Description of the datasets for experiments.}
% \begin{tabular}{lll}
% \hline
% Dataset & \# Classes & Usage \\ \hline
% CIFAR10 & 10& Victim / Suspect\\ %\hline
% MNIST & 10 & Victim / Suspect\\ %\hline
% Tiny-ImageNet & 100 & Victim / Suspect\\ %\hline
% ImageNet& 1000& Suspect \\ \hline
% \end{tabular}
% \label{tb:dataset}
% \end{table}
We trained victim models using three popularly used DNN architectures, ResNet18~\cite{ResNet18}, VGG16~\cite{vgg16}, and DenseNet161~\cite{DenseNet161}, on each of the victim datasets. Appendix~\ref{App:experiment_model} provides a summary of the information about those models.%The models using Tiny-ImageNet achieved relatively low accuracies, between 36--52\%.

% \begin{table}[h!]
% \centering
% \caption{Datasets, models, and parameters we used and \new{mean values along with the standard deviation values of baseline accuracy.}}
% \resizebox{2.6in}{!}{
% \begin{tabular}{llll}
% \hline
% Dataset & Architecture & \# Params & Accuracy (\%) \\ \hline
% & ResNet18&  11181642& \new{74.15 (0.37)}\\ \cline{2-4} 
% CIFAR10 & VGG16 &134301514& \new{82.62 (2.81)}\\ \cline{2-4} 
% & DenseNet161& 26494090& \new{70.80 (0.82)}\\ \hline
% & ResNet18 & 11181642& \new{99.47 (0.04)}\\ \cline{2-4}
% MNIST &  VGG16& 134301514 & \new{99.47 (0.04)}\\ \cline{2-4} 
% & DenseNet161& 26494090&  \new{99.26 (0.08)}\\ \hline
% & ResNet18 & 11181642&  \new{35.27 (0.63)}\\ \cline{2-4}
% Tiny-ImageNet &  VGG16& 134301514&  \new{39.46 (0.40)}\\ \cline{2-4} 
% & DenseNet161& 26494090&  \new{33.13 (2.18)}\\ \hline
% \end{tabular}}
% \label{tb:model_details}
% \end{table}

\subsection{Classifier Evaluation Settings}
\subheading{Training configuration.} We utilized the procedures described in Section~\ref{sec:meta_classifier} to construct a classifier capable of detecting whether a proprietary dataset has been used to build a model. In all experiments, we used a seed dataset consisting of 1888 randomly chosen images from CIFAR10, regardless of the datasets utilized for creating victim and suspect models. Subsequently, we generated 1888 adversarial DFT images for each victim model. These adversarial DFT images were split into two groups: 1600 images were used to train the classifier, and the remaining 288 images were used to determine the classifier's threshold and for testing purposes. We discuss the effects of training dataset size and dimensions in Appendix~\ref{Appendix:Effects of training dataset size and dimensions}.

%\subheading{Metrics.} \seonhye{We did not use these metrics anymore. How about remove this subheading?} We calculate three metrics: True Positive Rate (TPR), True Negative Rate (TNR), and Balanced Accuracy (BA). TPR means the ratio of correct answers (i.e., detecting the stolen model as ``Stolen'') when we test 288 adversarial samples of a \textit{stolen} model with a classifier. TNR means the ratio of correct answers (i.e., labelling benign model as ``Benign'') when testing 288 adversarial samples of a \textit{benign} model. BA is calculated as the average of TPR and TNR.
%The meta-classifier show -\% accuracy for the test set. We generate and observe adversarial DFTs for VGG16, ResNet18, and DenseNet161 models learned with ImageNet to measure false negative rates, resulting in 98.26\% false negative in average. In particular, the VGG model has 100 false negative rate. In order to observe whether detection is performed regardless of the architecture of the model, we trained the squeezezenet model not used for learning on CIFAR 100 and observed that it show 97.59\% accuracy. \sharif{Plese revisit all the numbers and datasets. Does not seem to sync now with latest status.}

\subheading{Effects of the threshold for the classifier.} 
In Section~\ref{sec:meta_classifier}, we discussed the process for determining the optimum threshold for a classifier using the validation dataset. We conducted experiments to investigate how the threshold value of the classifier impacts the performance of \sysname using nine models constructed from three datasets and three model architectures. We built a dedicated classifier for each dataset and evaluated its performance using balanced accuracy. Appendix~\ref{Appendix:thresholding} shows the relationship between each classifier's threshold value and balanced accuracy. We observed that increasing the threshold value for all classifiers tends to lead to a significant increase in balanced accuracy, reaching a certain point (i.e., 0.0025 for CIFAR10, 0.0005 for MNIST, and 0.003 for Tiny-ImageNet) before decreasing. We selected the threshold value for each classifier that maximises its balanced accuracy.

\subsection{Resilience against Data/Model Theft Attacks}
We evaluate the resilience of \sysname against the \new{eight} attack scenarios presented in Section~\ref{sec:threat_model}. \new{We repeated each attack scenario ten times. The mean values for Model Accuracy (Model Acc.) and \new{Theft image detection Rate} (\new{Theft Image Rate}) are presented, along with the standard deviation values in parentheses. The same format is used in the remaining tables throughout this paper.}
%In Section~\ref{sec:MAA}, we check the performance of three classifiers against MAA. In the other six attack scenarios, we use the classifier built to protect one of two datasets only. 
\subsubsection{Multi-Architecture Attack (MAA)}\label{sec:MAA}

We test \sysname against the MAA scenario. In this scenario, the attacker uses part of the stolen dataset to train a model with an architecture different from the original victim model. We chose three victim datasets -- CIFAR10, MNIST, and Tiny-ImageNet -- and trained each of them using three different model architectures (ResNet18, VGG16, and DenseNet161). We designate one dataset as the victim and assign the other datasets as benign. Additionally, the models trained on ImageNet~\cite{ImageNet} are used as benign models for MNIST and CIFAR10. 

\begin{figure}[!ht]
    \centering 
    \begin{tabular}{cc}
        \includegraphics[width=.93
        \linewidth]{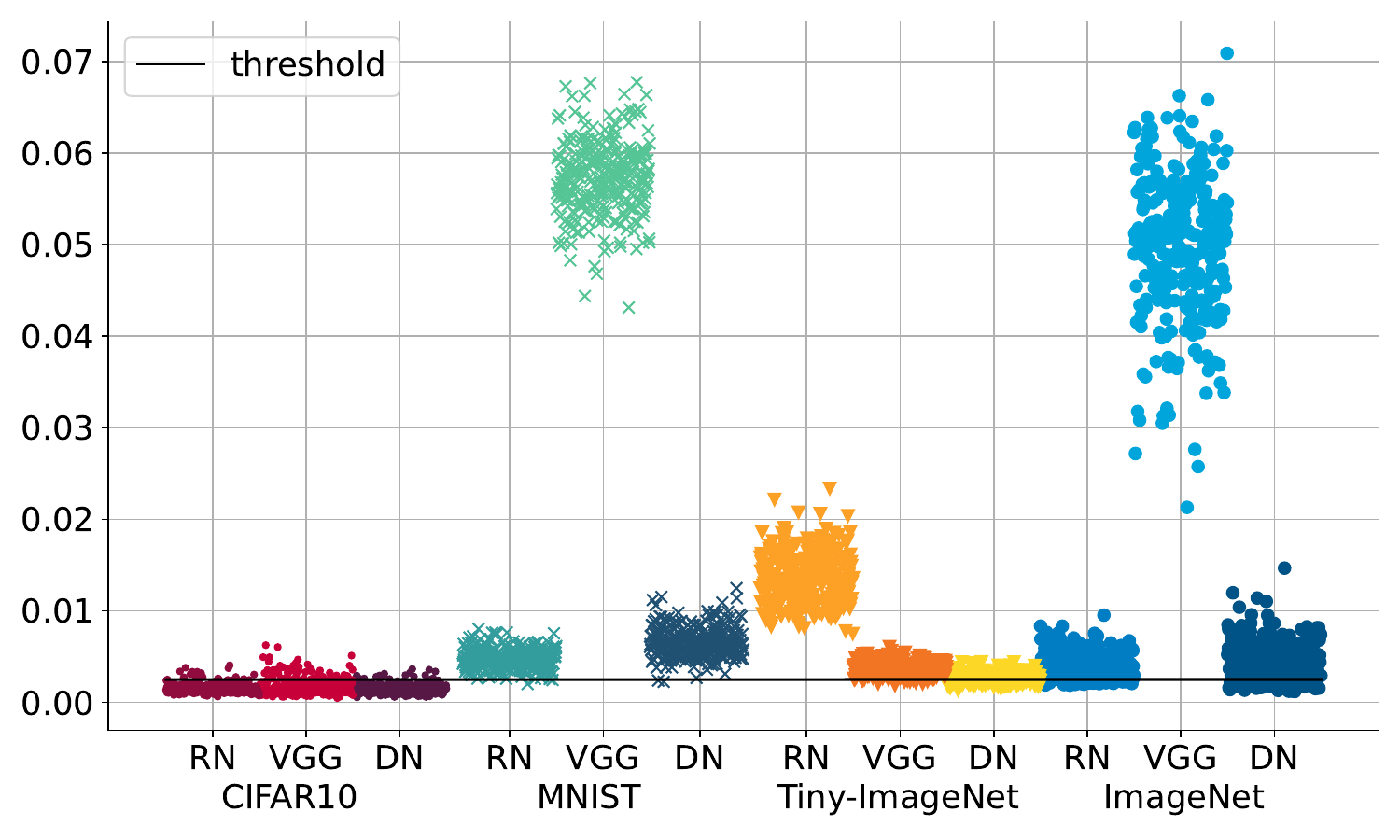}
    \end{tabular}
    \caption{Distribution of output scores produced by the classifier for CIFAR10 across 12 different models, each representing a combination of three architectures -- ResNet18 (RN), VGG16 (VGG), and DenseNet161 (DN) and four datasets -- CIFAR10, MNIST, Tiny-ImageNet, and ImageNet. The bold line represents the threshold chosen for \sysname.}
    \label{fig:threshold}
    \centering
\end{figure}

\subheading{Efficacy.} 
Table~\ref{tb:MAA_results} presents \sysname's performance against the MAA scenario. \sysname completely distinguishes attack cases from benign cases for all datasets. The results demonstrate that \sysname is highly effective against MAA under both stolen and benign scenarios. Figure~\ref{fig:threshold} visually demonstrates the effectiveness of a classifier used for \sysname. The figure displays the distribution of output scores produced by the classifier for CIFAR10 across 12 distinct models, each representing a combination of the three architectures (ResNet18, VGG16, and DenseNet161) and the four datasets (CIFAR10, MNIST, Tiny-ImageNet, and ImageNet). The bold line in the figure represents the chosen threshold for the classifier, which is trained on the CIFAR10. The classifier can successfully distinguish the DNN model using the CIFAR10 dataset from the models using the other datasets.

\begin{table}[!ht]
\centering
\caption{MAA results for CIFAR10, MNIST, Tiny-ImageNet classifiers with 12 suspect models, each representing a combination of three architectures -- ResNet18 (RN), VGG16 (VGG), and DenseNet161 (DN) and four datasets -- CIFAR10, MNIST, Tiny-ImageNet, and ImageNet. \new{The ``Copy Detection (\%)'' field values represent the successful copy detection rate. The same format is used in the remaining tables throughout the paper.}}
%\resizebox{\linewidth}{!}{
\resizebox{3.1in}{!}{
\begin{tabular}{c|c|c|c|c|c}
\hline
Victim & Suspect  & \begin{tabular}[c]{@{}c@{}}Ground \\ Truth\end{tabular} &\begin{tabular}[c]{@{}c@{}}Archi\\ tecture \end{tabular} &\begin{tabular}[c]{@{}c@{}} Theft Image\\ Rate \end{tabular}  &\begin{tabular}[c]{@{}c@{}} Copy \\ Detection (\%) \end{tabular}  \\ \hline
 &  &   & RN    & \new{93.01 (5.34)}     & \new{100}    \\ \cline{4-5}
 &  &   &VGG     & \new{84.65 (9.72)}   & \new{100}   \\ \cline{4-5}
 &  \multirow{-3}{*}{\begin{tabular}[c]{@{}c@{}} CIFAR10\end{tabular}} & \multirow{-3}{*}{\begin{tabular}[c]{@{}c@{}} Stolen\end{tabular}}   &  DN   & \new{94.55 (4.11)}& \new{100}\\ \cline{2-5}

 &  & & RN     & \new{9.17 (16.02)}   & \new{100}    \\ \cline{4-5}
 &  &   & VGG       & \new{0.0 (0.0)}    & \new{100}  \\ \cline{4-5}
 &  \multirow{-3}{*}{\begin{tabular}[c]{@{}c@{}}MNIST\end{tabular}} & \multirow{-3}{*}{\begin{tabular}[c]{@{}c@{}} Benign\end{tabular}}   &  DN   & \new{9.8 (11.48)}& \new{100} \\\cline{2-5}

 &  &   &RN    & \new{7.15 (8.01)}     & \new{100}   \\ \cline{4-5}
 &  &   &VGG        & \new{0.21 (0.52)}  &\new{100}   \\ \cline{4-5}
  &  \multirow{-3}{*}{\begin{tabular}[c]{@{}c@{}} Tiny- \\ ImageNet\end{tabular}} & \multirow{-3}{*}{\begin{tabular}[c]{@{}c@{}} Benign\end{tabular}}   &  DN   & \new{3.64 (4.33)}& \new{100}\\\cline{2-5}

 &  &   & RN     & \new{4.24 (7.25)}    & \new{100}    \\ \cline{4-5}
 &  &   &VGG       & \new{0.14 (0.23)}    & \new{100}   \\ \cline{4-5}
\multirow{-12}{*}{\begin{tabular}[c]{@{}c@{}} CIFAR10\end{tabular}}  &  \multirow{-3}{*}{\begin{tabular}[c]{@{}c@{}} ImageNet\end{tabular}} & \multirow{-3}{*}{\begin{tabular}[c]{@{}c@{}} Benign\end{tabular}}   &  DN   & \new{6.88 (12.94)}& \new{100}\\ \hline

 &  &   & RN    & \new{98.13 (3.51)}    & \new{100}    \\ \cline{4-5}
 &  &   &VGG     &\new{92.40 (9.62)}    & \new{100}   \\ \cline{4-5}
 &  \multirow{-3}{*}{\begin{tabular}[c]{@{}c@{}} MNIST\end{tabular}} & \multirow{-3}{*}{\begin{tabular}[c]{@{}c@{}} Stolen\end{tabular}}   &  DN   & \new{95.97 (7.13)}& \new{100}\\ \cline{2-5}

 &  &   & RN     & \new{4.69 (10.66)}   & \new{100}  \\ \cline{4-5}
 &  &   & VGG       & \new{0.0 (0.0)}   & \new{100}  \\ \cline{4-5}
 &  \multirow{-3}{*}{\begin{tabular}[c]{@{}c@{}}CIFAR10\end{tabular}} & \multirow{-3}{*}{\begin{tabular}[c]{@{}c@{}} Benign\end{tabular}}   &  DN   & \new{0.28 (0.72)}& \new{100}\\\cline{2-5}

 &  &   &RN    &\new{1.35 (2.14)}  &\new{100}    \\ \cline{4-5}
 &  &   &VGG        & \new{0.0 (0.0)} & \new{100}   \\ \cline{4-5}
  &  \multirow{-3}{*}{\begin{tabular}[c]{@{}c@{}} Tiny- \\ ImageNet\end{tabular}} & \multirow{-3}{*}{\begin{tabular}[c]{@{}c@{}} Benign\end{tabular}}   &  DN   & \new{0.14 (0.42)}&\new{100} \\\cline{2-5}

 &  &   & RN     & \new{5.14 (9.02)}  & \new{100}    \\ \cline{4-5}
 &  &   &VGG       &\new{10.03 (15.90)}   & \new{100}  \\ \cline{4-5}
\multirow{-12}{*}{\begin{tabular}[c]{@{}c@{}} MNIST\end{tabular}}  &  \multirow{-3}{*}{\begin{tabular}[c]{@{}c@{}} ImageNet\end{tabular}} & \multirow{-3}{*}{\begin{tabular}[c]{@{}c@{}} Benign\end{tabular}}   &  DN   &\new{6.27 (13.26)}& \new{100} \\ \hline

 &  &   & RN    & \new{92.71 (9.04)}    & \new{100}    \\ \cline{4-5}
 &  &   &VGG     &\new{72.61 (16.96)} & \new{100}   \\ \cline{4-5}
 &  \multirow{-3}{*}{\begin{tabular}[c]{@{}c@{}}Tiny-\\ImageNet\end{tabular}} & \multirow{-3}{*}{\begin{tabular}[c]{@{}c@{}} Stolen\end{tabular}}   &  DN   &\new{90.70 (7.07)}& \new{100}\\ \cline{2-5}

 &  &   & RN     &\new{6.01 (13.51)}  & \new{100}  \\ \cline{4-5}
 &  &   & VGG       &\new{11.35 (13.15)} & \new{100}   \\ \cline{4-5}
 &  \multirow{-3}{*}{\begin{tabular}[c]{@{}c@{}}CIFAR10\end{tabular}} & \multirow{-3}{*}{\begin{tabular}[c]{@{}c@{}} Benign\end{tabular}}   &  DN   & \new{4.20 (7.21)}& \new{100}\\\cline{2-5}

 &  &   & RN     &\new{1.32 (2.61)} & \new{100}     \\ \cline{4-5}
 &  &   &VGG       & \new{1.36 (3.52)}  & \new{100}   \\ \cline{4-5}
\multirow{-9}{*}{\begin{tabular}[c]{@{}c@{}} Tiny-\\ImageNet\end{tabular}}  &  \multirow{-3}{*}{\begin{tabular}[c]{@{}c@{}} MNIST\end{tabular}} & \multirow{-3}{*}{\begin{tabular}[c]{@{}c@{}} Benign\end{tabular}}   &  DN   & \new{7.43 (12.62)} & \new{100}\\ \hline

\end{tabular}}
\label{tb:MAA_results}
\end{table}

%\sharif{Explain the aim and the settings}

%The BA of the classifier of each of CIFAR10, MNIST, and Tiny-ImageNet shows high performance at 90.18\%, 94.95\%, and 93.60\%.

\observ{\sysname is effective in detecting DNN model theft attacks across various model architectures}

\subsubsection{Data Augmentation Attack (DAA)}\label{Section:daa}

In the DAA scenario, we assume that an attacker tries to obscure the use of a stolen dataset by augmenting it with new data. For example, if the CIFAR10 dataset is used as a victim dataset, we added five new classes from the CIFAR100 dataset to create a new dataset dubbed CIFAR15. The added classes include apples, bicycle, can, roses, and clock, which are distinct from any classes in the original CIFAR10 dataset. We train ResNet18 using the augmented CIFAR15 dataset as the attack case. The goal here is to evaluate whether \sysname can detect the usage of stolen CIFAR10 data, even when this data has been augmented with new classes. To test the effectiveness of \sysname at different stages of model training, we examine the performance of \sysname at five different epochs (20, 40, 60, 80, and 100), which can reveal if \sysname's detection ability changes as the model becomes more trained.
%Similarly, for Tiny-ImageNet as a victim dataset, we use the dataset of 110 classes by adding ten more classes of Tiny-ImageNet that were not used in the initial victim model training. We train ResNet18 on the new 110-class dataset as the attack case. For the MNIST, since MNIST has only 10 digits from zero to nine, we add the MNIST samples within each class using another dataset from the same domain named EMNIST.
%The mean accuracy value of these attack models is about 86.68\% (see Table~\ref{tb:daa_results} for complete accuracy results).

\subheading{Efficacy.} 
\new{Table~\ref{tb:daa_results} presents \sysname's performance against the DAA scenario. The \new{theft image detection rate} slightly increases as the training epoch of the model increases. While \sysname failed to detect one or two attack cases, \sysname is still effective against DAA for large training epochs.}

\begin{table}[h]
\caption{DAA results for CIFAR10 classifier.}
%\resizebox{\linewidth}{!}{
\resizebox{3.1in}{!}{
\begin{tabular}{c|c|c|c|c|c}
\hline
 Suspect  & \begin{tabular}[c]{@{}c@{}}Ground \\ Truth\end{tabular} &Epochs & \begin{tabular}[c]{@{}c@{}}Model \\ Acc.\end{tabular} & \begin{tabular}[c]{@{}c@{}}Theft Image\\ Rate \end{tabular} & \begin{tabular}[c]{@{}c@{}}Copy\\Detection (\%) \end{tabular}  \\ \hline
   & & 20 & \new{72.61 (0.39)}   & \new{77.12 (35.66)}     & \new{80}   \\ \cline{3-5}
   & & 40 & \new{72.56 (0.69)}     & \new{74.10 (33.34)}  & \new{80}   \\ \cline{3-5}
   & & 60 & \new{72.67 (0.59)}     & \new{72.01 (33.34)}    & \new{90}    \\ \cline{3-5}
   & & 80 & \new{72.5 (0.57)}     & \new{86.88 (16.50)}    & \new{90}   \\ \cline{3-5}
\multirow{-5}{*}{\begin{tabular}[c]{@{}c@{}} CIFAR10\end{tabular}}  &  \multirow{-5}{*}{\begin{tabular}[c]{@{}c@{}} Stolen\end{tabular}} & 100    &\new{72.67 (0.49)}      & \new{77.99 (26.51)}     & \new{90} \\ \hline
\end{tabular}}
\label{tb:daa_results}
\end{table}

\observ{\sysname is effective in detecting cases where a model has been trained on a mixed dataset of stolen data and new data, especially when using large training epochs}

\subsubsection{Same Architecture Attack (SAA)}

In the SAA scenario, an attacker trains the ResNet18 model on 10\%, 30\%, 50\%, 70\%, and 90\% of a victim dataset. We use CIFAR10 and MNIST, respectively, as victim datasets. We split the dataset uniformly, including an equal number of samples from every class. We examine the performance of \sysname at 10 epochs. %Since SAA experiment results vary depending on the random seed initialization value, we repeat these experiments three times and report the average results.

\subheading{Efficacy.} 
\new{Table~\ref{tb:MRA_results} presents the performance of \sysname against the SAA scenario. The experimental results indicate that for the MNIST dataset, \sysname can perfectly detect attacks when the portion of the stolen dataset used for training the model is 30\% or more. However, in the case of CIFAR10, only 80\% of the attacks were detected when the portion of the stolen dataset used for training was 70\%. This suggests that the likelihood of an attack being detected by \sysname increases with the amount of stolen data used for training. This observation is consistent with our intuition, as we expect a model to exhibit more characteristics of the stolen data when a larger portion is used for training.}% Likewise, \sysname can detect benign models trained on $\geq 70\%$ of MNIST dataset as benign with TNR of 64.24\%.  \\

\begin{table}[h]
\centering
\caption{SAA results for CIFAR10 and MNIST classifier when the different amount of dataset is used.}
%\resizebox{\linewidth}{!}{
\resizebox{3.1in}{!}{
\begin{tabular}{c|c|c|c|c|c}
\hline
 Victim &Suspect  & \begin{tabular}[c]{@{}c@{}}Ground \\ Truth\end{tabular} &\begin{tabular}[c]{@{}c@{}}Used Dataset \\ Size (\%)\end{tabular} & \begin{tabular}[c]{@{}c@{}}Theft Image\\ Rate \end{tabular} &\begin{tabular}[c]{@{}c@{}} Copy \\ Detection (\%) \end{tabular}    \\ \hline

  & & & 10    & \new{48.86 (24.56)}    &\new{40}   \\ \cline{4-5}
&  & & 30   &\new{57.05 (27.07)}   & \new{70}   \\ \cline{4-5}
 & & & 50     & \new{64.41 (29.02)}     & \new{60}  \\ \cline{4-5}
  & & & 70   &\new{80.03 (23.32)}     & \new{80}  \\ \cline{4-5}
\multirow{-5}{*}{\begin{tabular}[c]{@{}c@{}} CIFAR10\end{tabular}} &\multirow{-5}{*}{\begin{tabular}[c]{@{}c@{}} CIFAR10\end{tabular}}  &  \multirow{-5}{*}{\begin{tabular}[c]{@{}c@{}} Stolen\end{tabular}} & 90    & \new{79.10 (26.55)}       &\new{90}\\ \hline

 &  & & 10   & \new{65.56 (11.26)}    & \new{90}   \\ \cline{4-5}
 &  & & 30   & \new{80.35 (9.53)}    & \new{100}    \\ \cline{4-5}
 & & & 50   & \new{82.40 (8.59)}    & \new{100}   \\ \cline{4-5}
  & & & 70   & \new{93.61 (5.41)}    & \new{100}   \\ \cline{4-5}
\multirow{-5}{*}{\begin{tabular}[c]{@{}c@{}} MNIST\end{tabular}} &\multirow{-5}{*}{\begin{tabular}[c]{@{}c@{}}MNIST\end{tabular}}  &  \multirow{-5}{*}{\begin{tabular}[c]{@{}c@{}} Stolen\end{tabular}} & 90    &\new{95.14 (3.82)}    & \new{100}\\ \hline

%    & & 10   & 22.43     & No   \\ \cline{3-5}
%    & & 30      & 92.08    & Yes    \\ \cline{3-5}
%   & & 50    & 92.43   & Yes   \\ \cline{3-5}
%    & & 70     & 84.79    & Yes   \\ \cline{3-5}
% \multirow{-5}{*}{\begin{tabular}[c]{@{}c@{}}Tiny-\\ImageNet\end{tabular}}  &  \multirow{-5}{*}{\begin{tabular}[c]{@{}c@{}} Stolen\end{tabular}} & 90    &87.29    & Yes\\ \hline

\end{tabular}}
\label{tb:MRA_results}
\end{table}

\observ{\sysname is effective in detecting DNN model theft attacks when a significant portion (more than 70\%) of the stolen dataset is used for training the model}

\subsubsection{Transfer Learning Attack (TLA)}

%\sharif{Explain the aim and the settings}
In the TLA scenario, an attacker steals the victim's ResNet18 model trained on CIFAR10 and performs transfer learning with the MNIST dataset using a learning rate of 0.1 while freezing the lower 30\% layer. We examine the performance of \sysname at four different epochs \new{(2, 4, 6, and 8)}.

\subheading{Efficacy.} 
\new{Table~\ref{tb:tla_results} presents the performance of \sysname against the TLA scenario. \sysname failed to detect four attack cases when the attacker trained the model for only two epochs. However, \sysname was able to completely distinguish attack cases as stolen at 8 epochs.}

%the average attack probability and SD of the four Stolen TLA models with different training epochs is 97.31\% and (2.60\%). Also, the average TNR and SD of the four benign models with different training epochs is 99.57\% and (0.22\%). There is not much difference between training epochs, and overall it shows high accuracy in detecting transfer learning attacks. 

%Table~\ref{tb:tla_results} presents the performance of \sysname against the TLA scenario.

\begin{table}[h]
\caption{TLA results for CIFAR10 classifier.}
\centering
%\resizebox{\linewidth}{!}{
\resizebox{3.1in}{!}{
\begin{tabular}{c|c|c|c|c|c}
\hline
Suspect&\begin{tabular}[c]{@{}c@{}}Ground \\ Truth\end{tabular} & Epochs    & \begin{tabular}[c]{@{}c@{}} Model \\ Acc.\end{tabular} & \begin{tabular}[c]{@{}c@{}}Theft Image \\ Rate\end{tabular} & \begin{tabular}[c]{@{}c@{}} Copy \\ Detection (\%) \end{tabular}   \\ \hline
& &\new{2}    & \new{98.42 (0.21)} &  \new{54.31 (11.10)} & \new{60} \\ \cline{3-5}
& & \new{4}    & \new{98.65 (0.12)} & \new{62.09 (6.56)} & \new{100}  \\ \cline{3-5}
& & \new{6}    & \new{98.50 (0.27)} & \new{64.76 (12.08)}& \new{90}    \\ \cline{3-5}
\multirow{-4}{*}{\begin{tabular}[c]{@{}c@{}} CIFAR10 to\\ MNIST \end{tabular}} &  \multirow{-4}{*}{Stolen}& \new{8}   & \new{98.92 (0.21)} &  \new{69.93 (7.72)} & \new{100} \\ \hline
%  && 50  & 99.34  &   0.35 & \cellcolor[HTML]{BDF3BD}No    \\ \cline{3-5}
%  && \multicolumn{1}{c|}{100} & \multicolumn{1}{c|}{99.40} & \multicolumn{1}{c|}{0.0} &\cellcolor[HTML]{BDF3BD}No    \\ \cline{3-5}
%  && \multicolumn{1}{c|}{150} & \multicolumn{1}{c|}{99.40} & \multicolumn{1}{c|}{1.04} & \cellcolor[HTML]{BDF3BD}No    \\ \cline{3-5}
% \multirow{-4}{*}{\begin{tabular}[c]{@{}c@{}} Only \\ MNIST\end{tabular}}  
% & \multirow{-4}{*}{Benign} & \multicolumn{1}{c|}{200} & \multicolumn{1}{c|}{99.49} & \multicolumn{1}{c|}{0.35} 
% & \cellcolor[HTML]{BDF3BD}No  \\ \hline
\end{tabular}}
\label{tb:tla_results}
\end{table}

% \begin{table}[h]
% \caption{TLA results for CIFAR10 classifier with four positive models and four negative models. The copy field values below indicate the classification results (``Yes'' indicates stolen) and (``No'' indicates benign).}
% \centering
% \resizebox{2.7in}{!}{
% \begin{tabular}{c|c|c|c|c|c}
% \hline
% Suspect&\begin{tabular}[c]{@{}c@{}}Ground \\ Truth\end{tabular} & Epochs    & \begin{tabular}[c]{@{}c@{}} Model \\ Acc.\end{tabular} & \begin{tabular}[c]{@{}c@{}}Attack \\ Prob.\end{tabular} & Copy?   \\ \hline
% & & 20    & 99.34 &  96.53 & \cellcolor[HTML]{BDF3BD}Yes  \\ \cline{3-5}
% & & 40    & 99.40 &  93.40 & \cellcolor[HTML]{BDF3BD}Yes  \\ \cline{3-5}
% & & 60    & 99.40 &  100 & \cellcolor[HTML]{BDF3BD}Yes    \\ \cline{3-5}
% \multirow{-4}{*}{\begin{tabular}[c]{@{}c@{}} CIFAR10 to\\ MNIST \end{tabular}} &  \multirow{-4}{*}{Stolen}& 80    & 99.49 &  99.31 & \cellcolor[HTML]{BDF3BD}Yes \\ \hline
%  && 50  & 99.34  &   0.35 & \cellcolor[HTML]{BDF3BD}No    \\ \cline{3-5}
%  && \multicolumn{1}{c|}{100} & \multicolumn{1}{c|}{99.40} & \multicolumn{1}{c|}{0.0} &\cellcolor[HTML]{BDF3BD}No    \\ \cline{3-5}
%  && \multicolumn{1}{c|}{150} & \multicolumn{1}{c|}{99.40} & \multicolumn{1}{c|}{1.04} & \cellcolor[HTML]{BDF3BD}No    \\ \cline{3-5}
% \multirow{-4}{*}{\begin{tabular}[c]{@{}c@{}} Only \\ MNIST\end{tabular}}  
% & \multirow{-4}{*}{Benign} & \multicolumn{1}{c|}{200} & \multicolumn{1}{c|}{99.49} & \multicolumn{1}{c|}{0.35} 
% & \cellcolor[HTML]{BDF3BD}No  \\ \hline
% \end{tabular}}
% \label{tb:tla_results}
% \end{table}

\observ{\new{\sysname shows robust effectiveness in identifying cases where a stolen model has been further trained with new data through transfer learning when the model is fully trained}}

\begin{table}[h]
\caption{MFA results for CIFAR10 classifier.}
\centering
%\resizebox{\linewidth}{!}{
\resizebox{3.2in}{!}{
\begin{tabular}{c|c|c|c|c|c}
\hline
Suspect& \begin{tabular}[c]{@{}c@{}}Ground\\Truth\end{tabular} &\begin{tabular}[c]{@{}c@{}}Dataset\\Size\end{tabular} &  Model Acc. & Theft Image Rate &\begin{tabular}[c]{@{}c@{}} Copy \\ Detection (\%) \end{tabular}  \\ \hline
 && 500 (0.01\%)   & \new{74.67 (0.39)}  & \new{99.93 (0.14)}   &\new{100}  \\ \cline{3-5}
 && 1000 (0.02\%)   & \new{74.77 (0.22)}& \new{99.93 (0.14)} &\new{100}   \\ \cline{3-5}
\multirow{-3}{*}{\begin{tabular}[c]{@{}c@{}} CIFAR10 \end{tabular}} & \multirow{-3}{*}{\begin{tabular}[c]{@{}c@{}}Stolen\end{tabular}}  &2500 (0.05\%)   & \new{75.06 (0.04)}&\new{99.83 (0.23)}  & \new{100}  \\ \hline
& & 500 (0.01\%)   & \new{99.46 (0.0)}& \new{0.0 (0.0)} & \new{100}    \\ \cline{3-5}
 && 1000 (0.02\%)  & \new{99.48 (0.0)}& \new{0.0 (0.0)} & \new{100}     \\ \cline{3-5}
\multirow{-3}{*}{\begin{tabular}[c]{@{}c@{}}MNIST\end{tabular}}   &\multirow{-3}{*}{\begin{tabular}[c]{@{}c@{}}Benign\end{tabular}}   &2500 (0.05\%)   & \new{99.48 (0.0)}&\new{0.0 (0.0)} & \new{100}  \\ \hline
\end{tabular}}
\label{tb:MFA_results.}
\end{table}

\subsubsection{Model Fine-tuning Attack (MFA)}

In the MFA scenario, an attacker steals the ResNet18 model trained on CIFAR10 and the CIFAR10 dataset itself. The attacker then uses fine-tuning to train the stolen model using part of the CIFAR10 dataset, consisting of either 500, 1000, or 2500 samples, respectively, using the learning rate of 0.00005. For comparison, a benign model is trained on only the MNIST dataset using the same architecture. \new{We examine the performance of \sysname at 60 epochs.}

\subheading{Efficacy.} 
Table~\ref{tb:MFA_results.} presents the performance of \sysname against the MFA scenario. \sysname completely distinguishes all MFA cases from benign cases, irrespective of the used dataset size.

\observ{\sysname shows robust effectiveness in identifying cases where a stolen model has been further trained using a subset of the dataset used to train the stolen model}

\subsubsection{Model Pruning Attack (MPA)}

In the MPA scenario, an attacker steals the ResNet18 model trained on CIFAR10 and prunes 20\%, 40\%, and 60\% of the stolen model. Then the attacker fine-tunes the model for five epochs with a learning rate of 0.00005. For CIFAR10, we examine the performance of \sysname at 5 epochs. For comparison, a benign model is trained on only the MNIST dataset using the same architecture. For MNIST, we examine the performance of \sysname at 5 epochs.

\subheading{Efficacy.} 
Table~\ref{tb:mpa_results} presents the performance of \sysname against the MPA scenario. \sysname completely distinguishes all MPA instances from benign cases, regardless of the degree of pruning applied to the model. 

\begin{table}[!ht]
\caption{MPA results for CIFAR10 classifier with three positive models and three negative models.}
\resizebox{3.1in}{!}{
\begin{tabular}{c|c|c|c|c|c}
\hline
Suspect& \begin{tabular}[c]{@{}c@{}}Ground\\Truth\end{tabular} & \begin{tabular}[c]{@{}c@{}}Pruned\\ Percentage\end{tabular} & \begin{tabular}[c]{@{}c@{}}Model \\ Acc.\end{tabular} & \begin{tabular}[c]{@{}c@{}}Theft Image\\ Rate\end{tabular} & \begin{tabular}[c]{@{}c@{}} Copy \\ Detection (\%) \end{tabular}    \\ \hline
 &  & 20   & \new{63.09 (0.12)}  & \new{100 (0.0)}   &\new{100}  \\  \cline{3-5}
  & & 40   & \new{42.73 (0.07)} & \new{99.97 (0.11)} &\new{100}  \\  \cline{3-5}
\multirow{-3}{*}{\begin{tabular}[c]{@{}c@{}} CIFAR10 \end{tabular}} & \multirow{-3}{*}{\begin{tabular}[c]{@{}c@{}} Stolen \end{tabular}}&  60   & \new{21.44 (0.08)}  & \new{100 (0.0)} & \new{100} \\ \hline
  & & 20   &\new{98.53 (0.01)}  & \new{4.62 (7.22)}  & \new{100} \\  \cline{3-5}
 &  & 40   & \new{87.54 (0.11)}  & \new{1.25 (3.75)} & \new{100}  \\  \cline{3-5}
\multirow{-3}{*}{\begin{tabular}[c]{@{}c@{}}MNIST\end{tabular}}   & \multirow{-3}{*}{\begin{tabular}[c]{@{}c@{}} Benign\end{tabular}} & 60   & \new{43.18 (0.13)}  &\new{4.83 (14.48)} & \new{100}   \\ \hline
\end{tabular}}
\label{tb:mpa_results}
\end{table}

\observ{\sysname is robust against MPA regardless of the percentage of neurons pruned}

%the TPR of the pruned model is 96.53\%, 93.40\%, and 100\% as the percentage of the parameters pruned increases from 20\% to 60\%. \sysname provides a high detection TPR at 96.64\% (2.70\%) on average. On the other hand, the TNR is 99.31\%, 99.31\%, and 100\%, which is uniformly high regardless of how many parameters are pruned.% In the case of ROC AUC, the obtained results show high performance at 0.9445. \\

\subsubsection{Data Augmentation and Transfer Learning Attack (DATLA)}

%\sharif{Explain the aim and the settings}
In the DATLA scenario, an attacker creates the CIFAR15 dataset using the method explained in Section~\ref{Section:daa}. The attacker then uses a stolen ResNet18 model trained on the CIFAR10 dataset and fine-tunes it on the CIFAR15 dataset the attacker created. We examine the performance of \sysname at \new{five different epochs (20, 40, 60, 80, and 100).}
%For comparison, we also train a benign model on the MNIST dataset using the same architecture. We examine the performance of this model at the same three epochs. Similarly, for the MNIST dataset, an attacker creates a new dataset using the method explained in Section~\ref{Section:daa} and then uses a stolen ResNet18 model trained on the MNIST dataset and fine-tunes it on the new dataset the attacker created. In this case, we also train a benign model on the CIFAR10 dataset using the same architecture.

\subheading{Efficacy.} 
\new{Table~\ref{tb:datla_results} shows \sysname's performance against the DATLA scenario. \sysname failed to distinguish one or three cases when the training epoch is 20, 40, 60, and 80. However, \sysname successfully distinguished all attack cases when the training epoch is 100. This highlights the effectiveness of \sysname when using larger training epochs while also suggesting areas for potential improvement in handling cases with minimal training epochs.} 

\observ{\sysname is effective in detecting cases where a stolen model has been trained on a mixed dataset that combines stolen data with new data while suggesting areas for potential improvement in handling cases with minimal training epochs}

%The results show that \sysname can detect DATLA attack accurately with average and SD of the TPR of 96.64\% (2.70\%) and average and SD of the TNR of 99.54\% (0.33\%).
\begin{table}[t]
\caption{DATLA results for CIFAR10 classifier.}
%\resizebox{\linewidth}{!}{
\resizebox{3.1in}{!}{
\begin{tabular}{c|c|c|c|c|c}
\hline
 Suspect  & \begin{tabular}[c]{@{}c@{}}Ground \\ Truth\end{tabular} &Epochs & \begin{tabular}[c]{@{}c@{}}Model \\ Acc.\end{tabular} & \begin{tabular}[c]{@{}c@{}}Theft Image\\ Rate \end{tabular} & \begin{tabular}[c]{@{}c@{}} Copy \\ Detection (\%) \end{tabular}  \\ \hline

  & & 20 & \new{73.51 (0.74)}    & \new{84.62 (27.83)}    & \new{90}   \\ \cline{3-5}
  & & 40 & \new{73.53 (0.50)}    & \new{83.26 (24.61)}    & \new{90}   \\ \cline{3-5}
 & & 60 & \new{73.56 (0.47)}    & \new{82.92 (19.55)}    & \new{90}   \\ \cline{3-5}
  & & 80 & \new{73.63 (0.51)}    & \new{62.60 (36.40)}    & \new{70}   \\ \cline{3-5}
\multirow{-5}{*}{\begin{tabular}[c]{@{}c@{}} CIFAR10\end{tabular}}  &  \multirow{-5}{*}{\begin{tabular}[c]{@{}c@{}} Stolen\end{tabular}} & 100   & \new{73.24 (0.51)}      & \new{81.25 (11.36)}     & \new{100}  \\ \hline
\end{tabular}}
\label{tb:datla_results}
\end{table}

\subsubsection{\new{Transfer Learning with Pretrained model Attack (TLPA)}} \new{In the TLPA scenario, an attacker uses a pretrained model on a public dataset, such as MNIST. The attacker then steals the CIFAR10 dataset of 60,000 samples and uses transfer learning to fine-tune the pretrained model using the stolen dataset. Similar to the TLA scenario, the attacker performs transfer learning using a learning rate of 0.1 while freezing the lower 30\% layers. We examine the performance of \sysname at 10 epochs by varying the dataset size used to generate the pretrained model.}

\subheading{Efficacy.} 
Table~\ref{tb:tlpa_results} shows \sysname's performance against the TLPA scenario. The \new{theft Imagdetection rate} decreases as the dataset size used for the pre-trained model increases. \sysname can completely detect TLPA attacks when at least 30\% of the MNIST dataset is used for pretraining. However, if the dataset used for the pretrained model exceeds 50\%, \sysname may fail to detect some stolen cases.

\begin{table}[!ht]
\caption{\new{TLPA results for CIFAR10 classifier.}}
\resizebox{3.1in}{!}{
\begin{tabular}{c|c|c|c|c|c}
\hline
\new{Suspect}& \begin{tabular}[c]{@{}c@{}}\new{Ground}\\\new{Truth}\end{tabular} & \begin{tabular}[c]{@{}c@{}}\new{Used Dataset}\\ \new{Size (\%)}\end{tabular} & \begin{tabular}[c]{@{}c@{}}\new{Model} \\ \new{Acc.}\end{tabular} & \begin{tabular}[c]{@{}c@{}}\new{Theft Image} \\ \new{Rate}\end{tabular} & \begin{tabular}[c]{@{}c@{}} \new{Copy} \\ \new{Detection (\%)} \end{tabular}   \\ \hline
 &  & 10   & \new{67.66 (1.13)}  & \new{99.51 (1.46)}    &\new{100}  \\  \cline{3-5}
  & & 30   & \new{68.43 (0.63)}  &\new{99.58 (0.62) } &\new{100}  \\  \cline{3-5}
 & & 50   &\new{68.74 (0.86)}  & \new{77.99 (23.71)} &\new{90}   \\  \cline{3-5}
 & & 70   &\new{68.74 (0.86)} &\new{84.65 (13.94)} &\new{100}  \\  \cline{3-5}
\multirow{-5}{*}{\begin{tabular}[c]{@{}c@{}} MNIST \\to CIFAR10 \end{tabular}} & \multirow{-5}{*}{\begin{tabular}[c]{@{}c@{}} Stolen \end{tabular}}&  90   & \new{69.22 (0.64)}  &\new{85.28 (23.90)} & \new{80}  \\ \hline
\end{tabular}}
\label{tb:tlpa_results}
\end{table}

\observ{\new{\sysname shows robust effectiveness in identifying cases where a pretrained model has been further trained using a stolen dataset via transfer learning}}

\subsection{Comparison with Existing Fingerprinting}
\label{sec:Comparison with Existing Fingerprinting Techniques}

\subheading{Comparison settings.} We conduct experimental comparisons with \deepjudge~\cite{chen2021copy}, the state-of-the-art fingerprinting technique. \deepjudge generates four metrics for white-box evaluation and two metrics for black-box evaluation. It uses majority voting, where 3 out of 4 metrics have to produce values $<$ threshold to support the correct final judgment of being stolen. \deepjudge has been designed to provide architecture-dependent protection; namely, all model parameters need to be the same, including the number of classes. On the other hand, \sysname is designed to be architecture insensitive to enable the dataset intelligence to be tracked even when the model architecture is changed. Hence, \textit{\deepjudge is not able to detect MAA as stolen models owing to its design limitation}. Therefore, we compare \deepjudge with \sysname in the other \new{seven attacks} in addition to direct cloning. For the transfer learning attack, we observe that \deepjudge only considered the same number of classes between the original model and the transfer-learned model, which might not always be the case. Thus, we use the scripts released with \deepjudge and apply small modifications to run the data augmentation and transfer learning attacks when the number of classes is different. We set the target model of \deepjudge as the ResNet18 model trained on CIFAR10, MNIST, and Tiny-ImageNet. For the CIFAR10 dataset, we test with five attack models, using a ResNet18 model trained on MNIST and Tiny-ImageNet as the ``Benign'' cases. In other cases, we only check the \new{theft image detection rate} for two negative cases trained on other datasets. We used the optimal threshold for DeepJudge, which was optimized based on the testing results. These thresholds may be infeasible to achieve in practice, but it is ideal for comparison purposes. We only used four white-box metrics from DeepJudge.

\begin{table}[ht!]
\centering
\caption{Detection results of \deepjudge~\cite{chen2021copy}. \deepjudge uses a majority voting mechanism to determine whether a stolen model is being used. Specifically, \deepjudge classifies a model as stolen if at least three out of four metrics produce values $<$ threshold.}
\resizebox{3.1in}{!}{
%\resizebox{3.4in}{!}{
\begin{tabular}{clcccccc}
\hline
\multicolumn{1}{c|}{Victim}                          & \multicolumn{1}{c|}{\begin{tabular}[c]{@{}c@{}}Ground \\ Truth\end{tabular}}             & \multicolumn{1}{c|}{Suspect}       & \multicolumn{1}{c|}{Metric1} & \multicolumn{1}{c|}{Metric2}   &\multicolumn{1}{c|}{Metric3}& \multicolumn{1}{c|}{Metric4} & \multicolumn{1}{c}{\begin{tabular}[c]{@{}c@{}} Copy \\ Detection (\%) \end{tabular}}   \\ \hline\hline
%CIFAR10

\multicolumn{1}{c|}{}                                & \multicolumn{2}{c|}{\textbf{Threshold}}                          & \multicolumn{1}{c|}{\textbf{\new{0.0137}}}    & \multicolumn{1}{c|}{\textbf{\new{0.0542}}} & \multicolumn{1}{c|}{\textbf{\new{0.3544}}} & \multicolumn{1}{c|}{\textbf{\new{0.4255}}}   \\ \cline{2-8}
\multicolumn{1}{c|}{}                                & \multicolumn{1}{c|}{}                         & \multicolumn{1}{c|}{CIFAR10}       & \multicolumn{1}{c|}{\new{0.0 (0.0)}}  & \multicolumn{1}{c|}{\new{0.0 (0.0)}} & \multicolumn{1}{c|}{\new{0.0 (0.0)}} & \multicolumn{1}{c|}{\new{0.0 (0.0)}}   & \multicolumn{1}{c}{\new{100}}                       \\ \cline{3-7}
\multicolumn{1}{c|}{}                                & \multicolumn{1}{c|}{}                         & \multicolumn{1}{c|}{\begin{tabular}[c]{@{}c@{}}\new{CIFAR10} \\ \new{DAA}\end{tabular}} & \multicolumn{1}{c|}{\begin{tabular}[c]{@{}c@{}}\new{0.0055} \\ \new{(0.0032)}\end{tabular}}  & \multicolumn{1}{c|}{\begin{tabular}[c]{@{}c@{}}\new{0.0241} \\ \new{(0.0116)}\end{tabular}} & \multicolumn{1}{c|}{\begin{tabular}[c]{@{}c@{}}\new{0.5681} \\ \new{(0.3179)}\end{tabular}}  & \multicolumn{1}{c|}{\begin{tabular}[c]{@{}c@{}}\new{0.5771} \\ \new{(0.3503)}\end{tabular}}  & \multicolumn{1}{c}{\new{40}}    \\ \cline{3-7}
\multicolumn{1}{c|}{}                                & \multicolumn{1}{c|}{}                         & \multicolumn{1}{c|}{\begin{tabular}[c]{@{}c@{}}\new{CIFAR10} \\ \new{SAA}\end{tabular}} & \multicolumn{1}{c|}{\begin{tabular}[c]{@{}c@{}}\new{0.0050} \\ \new{(0.0023)}\end{tabular}}  & \multicolumn{1}{c|}{\begin{tabular}[c]{@{}c@{}}\new{0.0213} \\ \new{(0.0084)}\end{tabular}} & \multicolumn{1}{c|}{\begin{tabular}[c]{@{}c@{}}\new{1.0198} \\ \new{(0.3477)}\end{tabular}}  & \multicolumn{1}{c|}{\begin{tabular}[c]{@{}c@{}}\new{1.0123} \\ \new{(0.3324)}\end{tabular}}  & \multicolumn{1}{c}{\new{10}}                 \\ \cline{3-7}
\multicolumn{1}{c|}{}                                & \multicolumn{1}{c|}{}                         & \multicolumn{1}{c|}{\begin{tabular}[c]{@{}c@{}}\new{CIFAR10} \\ \new{TLA}\end{tabular}} & \multicolumn{1}{c|}{\begin{tabular}[c]{@{}c@{}}\new{0.0306} \\ \new{(0.0)}\end{tabular}}  & \multicolumn{1}{c|}{\begin{tabular}[c]{@{}c@{}}\new{0.1215} \\ \new{(0.0002)}\end{tabular}} & \multicolumn{1}{c|}{\begin{tabular}[c]{@{}c@{}}\new{0.3463} \\ \new{(0.0)}\end{tabular}}  & \multicolumn{1}{c|}{\begin{tabular}[c]{@{}c@{}}\new{0.3188} \\ \new{(0.0)}\end{tabular}}  & \multicolumn{1}{c}{\new{0}}                      \\ \cline{3-7}
\multicolumn{1}{c|}{}                                & \multicolumn{1}{c|}{}                         & \multicolumn{1}{c|}{\begin{tabular}[c]{@{}c@{}}\new{CIFAR10} \\ \new{MFA}\end{tabular}} & \multicolumn{1}{c|}{\begin{tabular}[c]{@{}c@{}}\new{0.0005} \\ \new{(0.0)}\end{tabular}}  & \multicolumn{1}{c|}{\begin{tabular}[c]{@{}c@{}}\new{0.0018} \\ \new{(0.0005)}\end{tabular}} & \multicolumn{1}{c|}{\begin{tabular}[c]{@{}c@{}}\new{0.0} \\ \new{(0.0)}\end{tabular}}  & \multicolumn{1}{c|}{\begin{tabular}[c]{@{}c@{}}\new{0.0} \\ \new{(0.0)}\end{tabular}}  & \multicolumn{1}{c}{\new{100}}                     \\ \cline{3-7}
\multicolumn{1}{c|}{}                                & \multicolumn{1}{c|}{} 
& \multicolumn{1}{c|}{\begin{tabular}[c]{@{}c@{}}\new{CIFAR10} \\ \new{MPA}\end{tabular}}           & \multicolumn{1}{c|}{\begin{tabular}[c]{@{}c@{}}\new{0.0108} \\ \new{(0.0001)}\end{tabular}}  & \multicolumn{1}{c|}{\begin{tabular}[c]{@{}c@{}}\new{0.0433} \\ \new{(0.0005)}\end{tabular}}   & \multicolumn{1}{c|}{\begin{tabular}[c]{@{}c@{}}\new{0.0830} \\ \new{(0.0)}\end{tabular}}& \multicolumn{1}{c|}{\begin{tabular}[c]{@{}c@{}}\new{0.0775} \\ \new{(0.0)}\end{tabular}}    &  \multicolumn{1}{c}{\new{100}}                     \\ \cline{3-7}
\multicolumn{1}{c|}{}                                & \multicolumn{1}{c|}{} 
& \multicolumn{1}{c|}{\begin{tabular}[c]{@{}c@{}}\new{CIFAR10} \\ \new{DATLA}\end{tabular}}           & \multicolumn{1}{c|}{\begin{tabular}[c]{@{}c@{}}\new{0.0071} \\ \new{(0.0027)}\end{tabular}}  & \multicolumn{1}{c|}{\begin{tabular}[c]{@{}c@{}}\new{0.0293} \\ \new{(0.010)}\end{tabular}}   & \multicolumn{1}{c|}{\begin{tabular}[c]{@{}c@{}}\new{0.1958} \\ \new{(0.0717)}\end{tabular}}& \multicolumn{1}{c|}{\begin{tabular}[c]{@{}c@{}}\new{0.1846} \\ \new{(0.0687)}\end{tabular}}    &  \multicolumn{1}{c}{\new{100}}                     \\ \cline{3-7}
\multicolumn{1}{c|}{}                             & \multicolumn{1}{c|}{\multirow{-14}{*}{Stolen}} & \multicolumn{1}{c|}{\begin{tabular}[c]{@{}c@{}}\new{CIFAR10} \\ \new{TLPA}\end{tabular}}      & \multicolumn{1}{c|}{\begin{tabular}[c]{@{}c@{}}\new{0.0101} \\ \new{(0.0022)}\end{tabular}}  & \multicolumn{1}{c|}{\begin{tabular}[c]{@{}c@{}}\new{0.0453} \\ \new{(0.0076)}\end{tabular}}   & \multicolumn{1}{c|}{\begin{tabular}[c]{@{}c@{}}\new{1.5621} \\ \new{(0.4715)}\end{tabular}}& \multicolumn{1}{c|}{\begin{tabular}[c]{@{}c@{}}\new{1.5654} \\ \new{(0.4720)}\end{tabular}}    & \multicolumn{1}{c}{\new{0}} \\ \cline{2-8}

\multicolumn{1}{c|}{}                                & \multicolumn{1}{c|}{}                         & \multicolumn{1}{c|}{MNIST}         & \multicolumn{1}{c|}{\begin{tabular}[c]{@{}c@{}}\new{0.0925 } \\ \new{(0.0331)}\end{tabular}}  & \multicolumn{1}{c|}{\begin{tabular}[c]{@{}c@{}}\new{0.3724} \\ \new{(0.1317)}\end{tabular}}   & \multicolumn{1}{c|}{\begin{tabular}[c]{@{}c@{}}\new{0.8535} \\ \new{(0.3554)}\end{tabular}} & \multicolumn{1}{c|}{\begin{tabular}[c]{@{}c@{}}\new{0.8597} \\ \new{(0.3503)}\end{tabular}}    & \multicolumn{1}{c}{\new{100}}        \\ \cline{3-7}

\multicolumn{1}{c|}{}                                & \multicolumn{1}{c|}{}                         & \multicolumn{1}{c|}{\begin{tabular}[c]{@{}c@{}}\new{MNIST} \\ \new{SAA}\end{tabular}} & \multicolumn{1}{c|}{\begin{tabular}[c]{@{}c@{}}\new{0.0806} \\ \new{(0.0328)}\end{tabular}}  & \multicolumn{1}{c|}{\begin{tabular}[c]{@{}c@{}}\new{0.2399} \\ \new{(0.1316)}\end{tabular}} & \multicolumn{1}{c|}{\begin{tabular}[c]{@{}c@{}}\new{1.1233} \\ \new{(0.5622)}\end{tabular}}  & \multicolumn{1}{c|}{\begin{tabular}[c]{@{}c@{}}\new{1.1528} \\ \new{(0.5460)}\end{tabular}}  & \multicolumn{1}{c}{\new{100}}   \\ \cline{3-7}

\multicolumn{1}{c|}{}                                & \multicolumn{1}{c|}{}                         & \multicolumn{1}{c|}{\begin{tabular}[c]{@{}c@{}}\new{MNIST} \\ \new{MFA}\end{tabular}} & \multicolumn{1}{c|}{\begin{tabular}[c]{@{}c@{}}\new{0.0846} \\ \new{(0.0)}\end{tabular}}  & \multicolumn{1}{c|}{\begin{tabular}[c]{@{}c@{}}\new{0.3408 } \\ \new{(0.0)}\end{tabular}} & \multicolumn{1}{c|}{\begin{tabular}[c]{@{}c@{}}\new{1.003} \\ \new{(0.0)}\end{tabular}}  & \multicolumn{1}{c|}{\begin{tabular}[c]{@{}c@{}}\new{1.032} \\ \new{(0.0)}\end{tabular}}  & \multicolumn{1}{c}{\new{100}}   \\ \cline{3-7}

\multicolumn{1}{c|}{}                                & \multicolumn{1}{c|}{}                         & \multicolumn{1}{c|}{\begin{tabular}[c]{@{}c@{}}\new{MNIST} \\ \new{MPA}\end{tabular}} & \multicolumn{1}{c|}{\begin{tabular}[c]{@{}c@{}}\new{0.664} \\ \new{(0.0)}\end{tabular}}  & \multicolumn{1}{c|}{\begin{tabular}[c]{@{}c@{}}\new{0.2705} \\ \new{(0.0004)}\end{tabular}} & \multicolumn{1}{c|}{\begin{tabular}[c]{@{}c@{}}\new{0.8244} \\ \new{(0.0)}\end{tabular}}  & \multicolumn{1}{c|}{\begin{tabular}[c]{@{}c@{}}\new{0.8810} \\ \new{(0.0)}\end{tabular}}  & \multicolumn{1}{c}{\new{100}}   \\ \cline{3-7}

\multicolumn{1}{c|}{\multirow{-25}{*}{CIFAR10}}       & \multicolumn{1}{c|}{\multirow{-9}{*}{Benign}} & \multicolumn{1}{c|}{\begin{tabular}[c]{@{}c@{}}\new{Tiny} \\ \new{ImageNet}\end{tabular}}  & \multicolumn{1}{c|}{\begin{tabular}[c]{@{}c@{}}\new{0.0164} \\ \new{(0.0050)}\end{tabular}}  & \multicolumn{1}{c|}{\begin{tabular}[c]{@{}c@{}}\new{0.0643} \\ \new{(0.0193)}\end{tabular}} & \multicolumn{1}{c|}{\begin{tabular}[c]{@{}c@{}}\new{1.2067} \\ \new{(0.4641)}\end{tabular}}  & \multicolumn{1}{c|}{\begin{tabular}[c]{@{}c@{}}\new{1.2083} \\ \new{(0.4415)}\end{tabular}}  & \multicolumn{1}{c}{\new{90}}   \\ \hline\hline

\multicolumn{1}{c|}{}            & \multicolumn{2}{c|}{\textbf{Threshold}}                  & \multicolumn{1}{c|}{\textbf{\new{0.0407}}}    & \multicolumn{1}{c|}{\textbf{\new{0.1846}}} & \multicolumn{1}{c|}{\textbf{\new{0.3022}}} & \multicolumn{1}{c|}{\textbf{\new{0.3094}}}   \\ \cline{2-8}
\multicolumn{1}{c|}{}    &  \multicolumn{1}{c|}{}   & \multicolumn{1}{c|}{MNIST}         & \multicolumn{1}{c|}{\new{0.0 (0.0)}}    & \multicolumn{1}{c|}{\new{0.0 (0.0)}} & \multicolumn{1}{c|}{\new{0.0 (0.0)}} & \multicolumn{1}{c|}{\new{0.0 (0.0)}}   & \multicolumn{1}{c}{\new{100}}     \\\cline{3-7}
\multicolumn{1}{c|}{}                                &  \multicolumn{1}{c|}{\multirow{-2}{*}{Stolen}}             & \multicolumn{1}{c|}{\begin{tabular}[c]{@{}c@{}}\new{MNIST} \\ \new{SAA}\end{tabular}}         & \multicolumn{1}{c|}{\begin{tabular}[c]{@{}c@{}}\new{0.042} \\ \new{(0.0279)}\end{tabular}}    & \multicolumn{1}{c|}{\begin{tabular}[c]{@{}c@{}}\new{0.17} \\ \new{(0.1106)}\end{tabular}} & \multicolumn{1}{c|}{\begin{tabular}[c]{@{}c@{}}\new{0.489} \\ \new{(0.3336)}\end{tabular}} & \multicolumn{1}{c|}{\begin{tabular}[c]{@{}c@{}}\new{0.491} \\ \new{(0.3338)}\end{tabular}}   & \multicolumn{1}{c}{\new{40}}      \\\cline{2-8}

\multicolumn{1}{c|}{}                                & \multicolumn{1}{c|}{}                         & \multicolumn{1}{c|}{CIFAR10}         & \multicolumn{1}{c|}{\begin{tabular}[c]{@{}c@{}}\new{0.0964} \\ \new{(0.0343)}\end{tabular}}    & \multicolumn{1}{c|}{\begin{tabular}[c]{@{}c@{}}\new{0.3860} \\ \new{(0.1371)}\end{tabular}}   & \multicolumn{1}{c|}{\begin{tabular}[c]{@{}c@{}}\new{0.8980} \\ \new{(0.5226)}\end{tabular}}  & \multicolumn{1}{c|}{\begin{tabular}[c]{@{}c@{}}\new{0.9060} \\ \new{((0.5413))}\end{tabular}}  & \multicolumn{1}{c}{\new{100}}                   \\ \cline{3-7}
\multicolumn{1}{c|}{\multirow{-7}{*}{MNIST}}         & \multicolumn{1}{c|}{\multirow{-3}{*}{Benign}} & \multicolumn{1}{c|}{\begin{tabular}[c]{@{}c@{}}\new{Tiny} \\ \new{ImageNet}\end{tabular}}      & \multicolumn{1}{c|}{\begin{tabular}[c]{@{}c@{}}\new{0.1027 } \\ \new{(0.3327)}\end{tabular}}    & \multicolumn{1}{c|}{\begin{tabular}[c]{@{}c@{}}\new{0.4111} \\ \new{(0.1329)}\end{tabular}} & \multicolumn{1}{c|}{\begin{tabular}[c]{@{}c@{}}\new{0.9823} \\ \new{(0.2940)}\end{tabular}}  & \multicolumn{1}{c|}{\begin{tabular}[c]{@{}c@{}}\new{0.9848} \\ \new{(0.2979)}\end{tabular}}    & \multicolumn{1}{c}{\new{100}} \\ \hline\hline

\multicolumn{1}{c|}{}            & \multicolumn{2}{c|}{\textbf{Threshold}}                  & \multicolumn{1}{c|}{\textbf{\new{0.0020}}}   & \multicolumn{1}{c|}{\textbf{\new{0.0095}}} & \multicolumn{1}{c|}{\textbf{\new{0.3660}}} & \multicolumn{1}{c|}{\textbf{\new{0.3660}}}          \\ \cline{2-8}
\multicolumn{1}{c|}{}    &  \multicolumn{1}{c|}{}   & \multicolumn{1}{c|}{\begin{tabular}[c]{@{}c@{}}\new{Tiny} \\ \new{ImageNet}\end{tabular}}         & \multicolumn{1}{c|}{\new{0.0 (0.0)}}    & \multicolumn{1}{c|}{\new{0.0 (0.0)}} & \multicolumn{1}{c|}{\new{0.0 (0.0)}} & \multicolumn{1}{c|}{\new{0.0 (0.0)}}   & \multicolumn{1}{c}{\new{100}}     \\\cline{3-7}
\multicolumn{1}{c|}{}                                &  \multicolumn{1}{c|}{\multirow{-2}{*}{Stolen}}             & \multicolumn{1}{c|}{\begin{tabular}[c]{@{}c@{}}\new{Tiny Image} \\ \new{-Net SAA}\end{tabular}}         & \multicolumn{1}{c|}{\begin{tabular}[c]{@{}c@{}}\new{0.0030} \\ \new{(0.0024)}\end{tabular}}    & \multicolumn{1}{c|}{\begin{tabular}[c]{@{}c@{}}\new{0.0131} \\ \new{(0.0091)}\end{tabular}} & \multicolumn{1}{c|}{\begin{tabular}[c]{@{}c@{}}\new{0.5111 } \\ \new{(0.3003)}\end{tabular}} & \multicolumn{1}{c|}{\begin{tabular}[c]{@{}c@{}}\new{0.5111} \\ \new{(0.3003)}\end{tabular}}   & \multicolumn{1}{c}{\new{30}}     \\\cline{2-8}

\multicolumn{1}{c|}{}                                & \multicolumn{1}{c|}{}                         & \multicolumn{1}{c|}{CIFAR10}       & \multicolumn{1}{c|}{\begin{tabular}[c]{@{}c@{}}\new{0.0052} \\ \new{(0.0019)}\end{tabular}}    & \multicolumn{1}{c|}{\begin{tabular}[c]{@{}c@{}}\new{0.0223 } \\ \new{(0.0069)}\end{tabular}} & \multicolumn{1}{c|}{\begin{tabular}[c]{@{}c@{}}\new{0.9156} \\ \new{(0.3672)}\end{tabular}} & \multicolumn{1}{c|}{\begin{tabular}[c]{@{}c@{}}\new{0.9156} \\ \new{(0.3672)}\end{tabular}}   & \multicolumn{1}{c}{\new{100}} \\ \cline{3-7}
\multicolumn{1}{c|}{\multirow{-7}{*}{\begin{tabular}[c]{@{}c@{}}Tiny-\\ ImageNet\end{tabular}}} & \multicolumn{1}{c|}{\multirow{-3}{*}{Benign}} & \multicolumn{1}{c|}{MNIST}         & \multicolumn{1}{c|}{\begin{tabular}[c]{@{}c@{}}\new{0.0047} \\ \new{(0.0013)}\end{tabular}}    & \multicolumn{1}{c|}{\begin{tabular}[c]{@{}c@{}}\new{0.0196} \\ \new{(0.0052)}\end{tabular}} & \multicolumn{1}{c|}{\begin{tabular}[c]{@{}c@{}}\new{0.8433} \\ \new{(0.3290)}\end{tabular}} & \multicolumn{1}{c|}{\begin{tabular}[c]{@{}c@{}}\new{0.8433} \\ \new{(0.3290)}\end{tabular}}   & \multicolumn{1}{c}{\new{100}} \\ \hline
                   
\end{tabular}}
\label{tb:comparison}
\end{table}

\subheading{Results.} 
%At first glance, \deepjudge seems to perform well in detecting model stealing attacks (see the ``Stolen'' cases in Table~\ref{tb:comparison}). However, \deepjudge was ineffective in detecting the ``Benign'' cases. For all ``Benign'' cases in the CIFAR10 and MNIST datasets as victim datasets, \deepjudge recognized them as ``Stolen.'' For the only Tiny-ImageNet dataset as a victim dataset, \deepjudge successfully recognized the benign models trained on CIFAR10 and MNIST as ``Benign.'' However, two metric values out of four were contained in the ranges of ``Stolen.''
Table~\ref{tb:comparison} presents the comparison results. For CIFAR, \deepjudge was ineffective in identifying most instances of theft against DAA (40\%), SAA (10\%), TLA (0\%), and TLPA (0\%). However, under the same settings, \sysname effectively detected those attacks (DAA (90\%), SAA (90\%), TLA (100\%), and TLPA (100\%)). For MNIST, the detection rate dropped to 40\% under the SAA scenario. Similarly, for Tiny-ImageNet, the detection rate dropped to 30\% under the SAA scenario.

\subsection{Robustness of \sysname with a Large-Sized Model Architecture}\label{sec:largemodel}

To evaluate the robustness of \sysname on large-sized model architectures, we used AlexNet~\cite{krizhevsky2017imagenet} as a representative architecture and Tiny-ImageNet as the victim dataset. We constructed a classifier with four victim models: AlexNet, ResNet18, VGG16, and DenseNet161, all trained on the Tiny-ImageNet dataset. We then evaluated the constructed classifier using twelve models, including the four victim and eight benign models trained on CIFAR10 or MNIST. As a result, \sysname remains effective in detecting adversarial images for large-scale model architectures such as AlexNet (see Appendix~\ref{app:large}). \new{We only conducted this experiment once due to the significant time required for experiments on AlexNet.}

% \begin{table}[!ht]
% \centering
% \caption{Performance of \sysname with AlexNet when the victim dataset is Tiny-ImageNet.}
% \resizebox{3.1in}{!}{
% \begin{tabular}{c|c|c|c|c}
% \hline
% Suspect & \begin{tabular}[c]{@{}c@{}} ground\\truth \end{tabular}& architecture & \begin{tabular}[c]{@{}c@{}} Theft image\\ Rate \end{tabular} & \begin{tabular}[c]{@{}c@{}} Copy \\Detection(\%) \end{tabular}\\ \hline
% \multirow{4}{*}{\begin{tabular}[c]{@{}c@{}} Tiny-\\ImageNet  \end{tabular}} & \multirow{4}{*}{stolen} & Alexnet &  95.83& \new{100}  \\ \cline{3-4}
%  &  & ResNet18 & 94.10 & \new{100} \\ \cline{3-4}
%  &  & VGG16 &  93.06 & \new{100}  \\ \cline{3-4}
%  &  & DenseNet161 & 94.79 &   \new{100}  \\ \cline{1-5}
% \multirow{4}{*}{\begin{tabular}[c]{@{}c@{}} CIFAR10 \end{tabular}} & \multirow{4}{*}{benign} & Alexnet &  100 &  \new{100}  \\ \cline{3-4}
%  &  & ResNet18 & 75.69 & \new{100}  \\ \cline{3-4}
%  &  & VGG16 &71.53 &  \new{100}  \\ \cline{3-4}
%  &  & DenseNet161 & 74.31 &   \new{100}  \\ \cline{1-5}
% \multirow{4}{*}{\begin{tabular}[c]{@{}c@{}} MNIST\end{tabular}} & \multirow{4}{*}{benign} & Alexnet &100 & \new{100}  \\ \cline{3-4}
%  &  & ResNet18 & 100 &   \new{100}   \\ \cline{3-4}
%  &  & VGG16 & 100 &   \new{100}  \\ \cline{3-4}
%  &  & DenseNet161 & 76.39 &  \new{100}   \\ \hline
% \end{tabular}}
% \label{tb:Alexnet}
% \end{table}

\subsection{\new{Ablation Study on DFT}}

\new{We conducted an ablation study to assess DFT's impact on \sysname's performance. We built a classifier identical to \sysname but without DFT, and tested its performance under the Multi-Architecture Attack scenario described in Section~\ref{sec:MAA}. We used four datasets (CIFAR10, MNIST, Tiny-ImageNet, and ImageNet) and three architectures (ResNet18, VGG16, and DenseNet161) for this evaluation.
Table~\ref{tb:without_DFT} indicates a significant degradation in accuracy when DFT is not applied. Without DFT, the classifier incorrectly classified all stolen cases as benign, except for one case in each CIFAR10 and MNIST dataset when the DenseNet161 architecture was used. Conversely, as shown in Table~\ref{tb:MAA_results}, \sysname could correctly detect benign and stolen cases with DFT. These findings underscore the critical role of DFT in maintaining the efficacy of \sysname.}

\begin{table}[!ht]
\centering
\caption{\new{Ablation study on DFT results.}}
%\resizebox{\linewidth}{!}{
\resizebox{3.1in}{!}{
\begin{tabular}{c|c|c|c|c|c}
\hline
Victim & Suspect  & \begin{tabular}[c]{@{}c@{}}Ground \\ Truth\end{tabular} &\begin{tabular}[c]{@{}c@{}}Archi\\ tecture \end{tabular} &Theft Image Rate  &\begin{tabular}[c]{@{}c@{}} Copy \\ Detection (\%) \end{tabular}  \\ \hline
 &  &   & RN    & \new{4.62 (11.82)}     & \new{0}    \\ \cline{4-5}
 &  &   &VGG     & \new{5.35 (13.98)}   & \new{0}   \\ \cline{4-5}
 &  \multirow{-3}{*}{\begin{tabular}[c]{@{}c@{}} CIFAR10\end{tabular}} & \multirow{-3}{*}{\begin{tabular}[c]{@{}c@{}} Stolen\end{tabular}}   &  DN   & \new{11.25 (22.62)}&  \new{10}\\ \cline{2-5}

 &  &   & RN     & \new{4.55 (10.19)}   &  \new{100}    \\ \cline{4-5}
 &  &   & VGG       & \new{0.0 (0.0)}    &  \new{100}  \\ \cline{4-5}
 &  \multirow{-3}{*}{\begin{tabular}[c]{@{}c@{}}MNIST\end{tabular}} & \multirow{-3}{*}{\begin{tabular}[c]{@{}c@{}} Benign\end{tabular}}   &  DN   & \new{1.39 (4.17)}&  \new{100} \\\cline{2-5}

 &  &   &RN    & \new{0.0 (0.0)}     &  \new{100}   \\ \cline{4-5}
 &  &   &VGG        & \new{0.14 (0.42)}  & \new{100}   \\ \cline{4-5}
  &  \multirow{-3}{*}{\begin{tabular}[c]{@{}c@{}} Tiny- \\ ImageNet\end{tabular}} & \multirow{-3}{*}{\begin{tabular}[c]{@{}c@{}} Benign\end{tabular}}   &  DN   & \new{1.04 (2.09)}&  \new{100}\\\cline{2-5}

 &  &   & RN     & \new{15.97 (25.64)}  &  \new{90}    \\ \cline{4-5}
 &  &   &VGG       &\new{8.09 (17.51)}   &  \new{90}  \\ \cline{4-5}
\multirow{-12}{*}{\begin{tabular}[c]{@{}c@{}} CIFAR10\end{tabular}}  &  \multirow{-3}{*}{\begin{tabular}[c]{@{}c@{}} ImageNet\end{tabular}} & \multirow{-3}{*}{\begin{tabular}[c]{@{}c@{}} Benign\end{tabular}}   &  DN   & \new{12.78 (23.02)}&  \new{90}\\ \hline

 &  &   & RN    & \new{4.51 (10.05)}    &  \new{0}    \\ \cline{4-5}
 &  &   &VGG     &\new{0.0 (0.0)}    &  \new{0}   \\ \cline{4-5}
 &  \multirow{-3}{*}{\begin{tabular}[c]{@{}c@{}} MNIST\end{tabular}} & \multirow{-3}{*}{\begin{tabular}[c]{@{}c@{}} Stolen\end{tabular}}   &  DN   & \new{11.08 (29.77)}&  \new{10}\\ \cline{2-5}

 &  &   & RN     & \new{0.42 (0.95)}   &  \new{100}  \\ \cline{4-5}
 &  &   & VGG       & \new{0.17 (0.42)}   &  \new{100}  \\ \cline{4-5}
 &  \multirow{-3}{*}{\begin{tabular}[c]{@{}c@{}}CIFAR10\end{tabular}} & \multirow{-3}{*}{\begin{tabular}[c]{@{}c@{}} Benign\end{tabular}}   &  DN   & \new{2.40 (5.99)}&  \new{100}\\\cline{2-5}

 &  &   &RN    &\new{0.035 (0.105)}  & \new{100}    \\ \cline{4-5}
 &  &   &VGG        & \new{0.0 (0.0)} &  \new{100}   \\ \cline{4-5}
  &  \multirow{-3}{*}{\begin{tabular}[c]{@{}c@{}} Tiny- \\ ImageNet\end{tabular}} & \multirow{-3}{*}{\begin{tabular}[c]{@{}c@{}} Benign\end{tabular}}   &  DN   & \new{0.35 (0.93)}&  \new{100} \\\cline{2-5}

 &  &   & RN     & \new{1.15 (2.64)}    &  \new{100}    \\ \cline{4-5}
 &  &   &VGG       & \new{0.35 (0.71)}    &  \new{100}   \\ \cline{4-5}
\multirow{-12}{*}{\begin{tabular}[c]{@{}c@{}} MNIST\end{tabular}}  &  \multirow{-3}{*}{\begin{tabular}[c]{@{}c@{}} ImageNet\end{tabular}} & \multirow{-3}{*}{\begin{tabular}[c]{@{}c@{}} Benign\end{tabular}}   &  DN   &\new{0.56 (1.35)}&  \new{100} \\ \hline

 &  &   & RN    & \new{0.07 (0.14)}    &  \new{0}    \\ \cline{4-5}
 &  &   &VGG     &\new{0.0 (0.0)} &  \new{0}   \\ \cline{4-5}
 &  \multirow{-3}{*}{\begin{tabular}[c]{@{}c@{}}Tiny-\\ImageNet\end{tabular}} & \multirow{-3}{*}{\begin{tabular}[c]{@{}c@{}} Stolen\end{tabular}}   &  DN   &\new{0.10 (0.22)}&  \new{0}\\ \cline{2-5}

 &  &   & RN     &\new{0.07 (0.14)}  &  \new{100}  \\ \cline{4-5}
 &  &   & VGG       &\new{0.04 (0.11)} &  \new{100}   \\ \cline{4-5}
 &  \multirow{-3}{*}{\begin{tabular}[c]{@{}c@{}}CIFAR10\end{tabular}} & \multirow{-3}{*}{\begin{tabular}[c]{@{}c@{}} Benign\end{tabular}}   &  DN   & \new{0.04 (0.11)}&  \new{100}\\\cline{2-5}

 &  &   & RN     &\new{0.17 (0.32)} &  \new{100}     \\ \cline{4-5}
 &  &   &VGG       & \new{0.35 (0.11)}  &  \new{100}   \\ \cline{4-5}
\multirow{-9}{*}{\begin{tabular}[c]{@{}c@{}} Tiny-\\ImageNet\end{tabular}}  &  \multirow{-3}{*}{\begin{tabular}[c]{@{}c@{}} MNIST\end{tabular}} & \multirow{-3}{*}{\begin{tabular}[c]{@{}c@{}} Benign\end{tabular}}   &  DN   & \new{0.0 (0.0)} &  \new{100}\\ \hline

\end{tabular}}
\label{tb:without_DFT}
\end{table}

\subsection{Detection Latency Results}
We assessed the detection latency of \sysname by analyzing the time taken by the classifier to identify a suspect model. The process consisted of three steps, as shown in Table~\ref{tb:complexity}.

\begin{table}[!ht]
\centering
\caption{Mean time taken for each step in \sysname.}
\resizebox{2.3in}{!}{
\begin{tabular}{c|c|c}
\hline
Step & Task                       & Time (Sec)         \\ \hline
1    & Adversarial DFT Generation & 668.0 \\ \hline
2   & Classifier Training   & 766.8              \\ \hline
3    & Suspect Model Verification & 9.78 \\ \hline
\end{tabular}}
\label{tb:complexity}
\end{table}

The first step involved the generation of adversarial images in the DFT domain. This took approximately 668 seconds, equivalent to 0.3 seconds per image. The second step comprised training the classifier, taking about 766.8 seconds. It is important to note that the first two steps are one-off tasks, meaning they are not repeated for each verification. The final step was verifying the suspect model, taking around 9.78 seconds, equivalent to 0.03 seconds per image. The entire process, including generation, training, and verification, took approximately 1444.51 seconds. These latency results are comparable to those of \deepjudge, which took 1937.79 seconds.

We plan to optimize the verification process by reducing the number of adversarial images tested. We aim to use only a few images for verification. This approach should enable \sysname to verify a suspect model within 1 second while still maintaining reasonable detection accuracy.
%\seonhye{I will add more details about how pick the seed images}
%-------------------------------------------------------------------------------
\section{Discussion}
\label{sec:Discussion}
%-------------------------------------------------------------------------------
%\subsection{Sensitivity Evaluation}

\subheading{Usefulness of \sysname.}
\sysname exhibits usefulness in various aspects. First, to ensure robustness, \sysname effectively detected \new{DNN model theft} attacks under various conditions. For example, \sysname was able to detect attacks on models trained on different datasets, attacked with different adversarial attack methods, and even attacked by models trained with a small subset of the victim dataset. Second, to ensure fidelity, \sysname has zero impact on model accuracy as it uses a fingerprinting technique rather than invasive watermarking. This means that \sysname can be used to protect models without sacrificing their performance. Third, to ensure efficacy, \sysname can distinguish most of the attack cases against benign models, except for some extreme cases (i.e., trained with small epochs or trained with less than 70\% of the victim dataset). This means that \sysname is a highly effective tool for detecting \new{DNN model theft} attacks. Overall, \sysname is a robust, reliable, and effective tool for protecting machine learning models from theft.

\subheading{Robustness against unseen architectures.}
\sysname was specifically designed to detect DNN model theft attacks across various architectures. However, our experiments suggest that \sysname may not be effective in detecting models trained using completely new or unseen architectures (see Table~\ref{tb:unseen}).

\begin{table}[h]
\centering
\caption{\new{Performance of \sysname with unseen architectures when the victim dataset is CIFAR10.}}
\resizebox{3.1in}{!}{
\begin{tabular}{c|c|c|c|c}
\hline
Architecture & dataset & \begin{tabular}[c]{@{}c@{}} ground \\ truth \end{tabular} & \begin{tabular}[c]{@{}c@{}} Theft Image\\ Rate \end{tabular} & \begin{tabular}[c]{@{}c@{}} Copy \\Detection(\%) \end{tabular}\\ \hline
\multirow{3}{*}{ResNet101} & CIFA10 & stolen  & 0.0 &0 \\ \cline{2-4}
 & MNIST &benign & 0.0  &100  \\ \cline{2-4}
 & Tiny-ImageNet & benign&  0.0 & 100 \\ \hline
\multirow{3}{*}{SqueezeNet} & CIFA10 & stolen  & 0.0 & 0\\ \cline{2-4}
 & MNIST &benign  &0.0  & 100  \\ \cline{2-4}
 & Tiny-ImageNet & benign & 0.0  & 100 \\ \hline
\end{tabular}}
\label{tb:unseen}
\end{table}

To evaluate the effectiveness of \sysname against models with an unseen architecture, we built a classifier using the CIFAR10 dataset as the victim dataset. We then tested the classifier's robustness against six different models, three of which were trained using ResNet101 and the other three using SqueezeNet~\cite{squeezenet}. The evaluation results are presented in Table~\ref{tb:unseen}.

Unfortunately, our experiments revealed that \sysname was ineffective in detecting the use of a stolen dataset (CIFAR10) for both ResNet101 and SqueezeNet models. This is because the generated adversarial images are influenced not only by the dataset used but also by the underlying model architecture. Based on these findings, we conclude that \sysname may not be effective in detecting models trained using completely new or unseen architectures. 

To overcome this limitation, we need to build a more diverse and comprehensive classifier that includes a broader range of architectures. As demonstrated in Section~\ref{sec:largemodel}, our experimental results indicate that \sysname effectively detects dataset theft using AlexNet when the classifier is constructed using the same architecture (i.e., AlexNet). However, the effectiveness of \sysname declines when the attacker's model is not included in the classifier construction phase. Nonetheless, it is worth noting that \sysname is currently the only fingerprinting scheme that can operate across multiple architectures.

\subheading{\new{Robustness against watermark removal attacks.}} \new{Various watermark removal attacks have been introduced to evaluate the robustness of DNN watermarking techniques~\cite{Aiken2020Neural}. Traditional watermark removal attacks are ineffective because DeepTaster does not rely on specialized watermark-triggered input samples. However, a recent study by Guo \textit{et al.}~\cite{guo2020fine} introduced a new watermark removal attack technique that uses preprocessing. They discovered that watermark samples are less robust than normal samples and designed a preprocessing function to compromise the watermark verification output without affecting the normal output. This approach employs a series of transformations, such as scaling, embedding random imperceptible patterns, and spatial-level transformations, to effectively disable watermark-triggered input samples while maintaining the model's accuracy. The preprocessing methods used in this scheme could make it more challenging to find effective adversarial examples for DeepTaster. We plan to explore this possibility in future work.}

%\newline\newline
%\subsubsection{Adaptive strategies}
%\subsubsection{Adaptive threshold}
\subheading{\new{Robustness against adversarial training.}} \new{To evaluate the impact of adversarial training on \sysname's effectiveness, we conducted experiments using the ResNet18 architecture and the CIFAR10 dataset. We used an adversarial training method~\cite{andriushchenko2020understanding} than the method introduced by Madry et al.~\cite{PGD}. While Madry et al.'s method was originally designed to improve robustness, the method we used focuses on maintaining model accuracy when training on adversarial examples. As a result, our model maintains its accuracy after adversarial training. In this attack scenario, an attacker steals the CIFAR10 dataset and generates adversarial examples using the FGSM attack with an epsilon of 0.001. Our method, on the other hand, aims to maximize model accuracy. The attacker then trains the ResNet18 model on these adversarial examples, which comprise 1\%, 3\%, 5\%, 7\%, and 9\% of the original CIFAR10 dataset, respectively. We assessed \sysname's performance over 10 epochs, repeating the same experiments 10 times for each configuration to avoid bias.}

\new{Table~\ref{tb:ala_results} presents the results. We found that \sysname's performance decreased as the proportion of adversarial examples in the training data increased. Even when only 1\% of adversarial examples were used, \sysname failed once out of 10 attempts. When 5\% or more of the training data consisted of adversarial examples, \sysname failed as many as four times. 
This result is not surprising, as \sysname relies on the characteristics of adversarial examples to determine the model's decision boundary. When new adversarial examples are introduced during training, the existing examples used for fingerprinting struggle to accurately reflect the model's decision boundary. To overcome this limitation, future research could explore pretraining victim models with adversarial training before applying \sysname.}

% \observ{\new{\sysname is effective in detecting adversarial learning attacks when a small size (about 3\%) of dataset is used for trained the model.}}

\begin{table}[!ht]
\caption{Performance of \sysname against adversarial training.}
\resizebox{3.1in}{!}{
\begin{tabular}{c|c|c|c|c|c}
\hline
Suspect& \begin{tabular}[c]{@{}c@{}}Ground\\Truth\end{tabular} & \begin{tabular}[c]{@{}c@{}}Used Dataset\\ Size (\%)\end{tabular} & \begin{tabular}[c]{@{}c@{}}Model \\ Acc.\end{tabular} & \begin{tabular}[c]{@{}c@{}}Theft Image \\ Rate\end{tabular} & \begin{tabular}[c]{@{}c@{}} Copy \\ Detection (\%) \end{tabular}    \\ \hline
 &  & 1   & 70.45 (0.96)  & 75.04 (16.70)    &\new{90}  \\  \cline{3-5}
  & & 3   & 70.15 (1.02)  & 66.74 (30.32)  &\new{80}  \\  \cline{3-5}
 & & 5   &70.52 (0.88)  & 56.67 (26.60)  &\new{60} \\  \cline{3-5}
 & & 7   & 70.53 (0.59)  &54.83 (23.21) &\new{60}  \\  \cline{3-5}
\multirow{-5}{*}{\begin{tabular}[c]{@{}c@{}} CIFAR10 \end{tabular}} & \multirow{-5}{*}{\begin{tabular}[c]{@{}c@{}} Stolen \end{tabular}}&  9   & 70.24 (0.64)  & 43.20 (31.46)  & \new{60} \\ \hline
\end{tabular}}
\label{tb:ala_results}
\end{table}

%\sysname is similar to membership inference attacks in that it infers that a target model contains samples from  dataset. Hence, Membership inference attacks might also be used as a fingerprinting technique. However, those attacks generally require a larger number of samples than \sysname.General membership inference attacks differ in that the goal is to determine whether a particular data has been used for target model training, and we aim to distinguish the training dataset. But the membership reference attack allows attackers to find the training dataset by attacking with several data in the same dataset. We conducted a comparative experiment with the state-of-art model membership inference attack, TrajectoryMIA~\cite{Yiyong2022Membership}. 

%\seonhye{I added this subheading for comparison to membership inference attack. After finished experiments, I will fill the missing words.}

\subheading{Comparison to membership inference attacks.} \sysname and membership inference attacks are both methods to infer whether a target model contains samples from a dataset. Therefore, conventional membership inference attacks could potentially be used as a model fingerprinting technique. 

To discuss the effectiveness of \sysname in detecting stolen datasets against membership inference attacks, we conducted experiments using TrajectoryMIA~\cite{Yiyong2022Membership}, a state-of-the-art membership inference attack method. TrajectoryMIA generates $k$ different distillation models based on a target model and its shadow model, both trained on different datasets, during $k$ epochs. The trajectory losses of a sample through each distillation model are computed to determine whether the sample is a member of the target model. TrajectoryMIA's membership inference results can be used for model fingerprinting. If a sample from a victim dataset is a member of a suspect model, we can conclude that the suspect model was built on the victim dataset. To evaluate the performance of TrajectoryMIA, we used a VGG16 model trained on CIFAR10 as the suspect model. We used the same VGG16 model for its shadow and distillation models with the settings presented in~\cite{Yiyong2022Membership}. The accuracy achieved by TrajectoryMIA on all CIFAR10 samples used as testing data was 60.52\%. To compare the effectiveness of \sysname with TrajectoryMIA, we built a classifier of \sysname consisting of three architectures: ResNet18, VGG16, and DenseNet161. We evaluated the performance of \sysname using 288 adversarial images as testing samples for the victim model (i.e., VGG16 trained on CIFAR10), and it achieved a 98.26\% accuracy. Although this comparison may not be entirely fair, we argue that utilizing a membership inference method in a straightforward manner might be insufficient for model fingerprinting. Moreover, \sysname is expected to be more efficient than TrajectoryMIA when testing multiple suspect models simultaneously since TrajectoryMIA requires creating several distillation models for each suspect model, whereas \sysname requires building a classifier only once for all suspect models.\looseness=-1

\subheading{Effects of the type of adversarial examples.}
The type of adversarial example can affect the performance of \sysname. To investigate this, we experimented with two popular attacks: PGD~\cite{madry2017towards} and FGSM~\cite{FGSM}. We compared the accuracy of the classifier on CIFAR10 using suspect models trained on CIFAR10, MNIST, and Tiny-ImageNet. \sysname using FGSM was about 7.73\% more accurate than \sysname using PGD. Therefore, we recommend using the FGSM attack to build \sysname. Finding the most effective adversarial DFT images for a given suspect model is a challenging problem. As part of our future work, we plan to develop an algorithm to identify effective adversarial examples for \sysname.\looseness=-1

\subheading{\new{On the high standard deviation of theft image detection rate:}} \new{We observed a high standard deviation in the theft image detection rate, particularly for models trained from scratch compared to pre-trained models. This variance was prominent in specific attack scenarios: DAA (Table~\ref{tb:daa_results}), SAA (Table~\ref{tb:MRA_results}), TLPA (Table~\ref{tb:tlpa_results}), and adversarial training (Table~\ref{tb:ala_results}). An exception is DATLA (Table~\ref{tb:datla_results}), which requires transfer learning on a dataset with different classes, demanding more parameter adjustments. This may explain why \sysname can occasionally fail to detect theft images in these scenarios, resulting in an inflated standard deviation. However, the standard deviation is not as elevated when considering only successful instances of \sysname. For example, DAA reduced from 28.05\% to 13.12\%, SAA from 26.10\% to 12.65\%, DATLA from 23.95\% to 11.5\%, TLPA from 12.73\% to 7.32\%, and adversarial training from 25.66\% to 12.96\%. The high standard deviation can be attributed to the small sample size (10 experiments) and the inclusion of significantly lower detection rate outliers from failed cases. These findings indicate that the performance of \sysname can be inconsistent in specific attack scenarios. To address this limitation in the future, we may consider adopting more customized and optimized detection decision criteria than the fixed 50\% threshold.}

\section{Related Work}
% %-------------------------------------------------------------------------------
% \subsection{DNNs fingerprinting}
% IPGuard \cite{IPGuard}
% DeepJudge \cite{deepjudge}

% \subsection{Dataset Protection}

% \subsection{DNNs IP attacks}
\subheading{DNN watermarking.} 
The first stream of related work uses watermarking to protect the copyright of DNN models~\cite{uchida2017embedding,adi2018turning,zhang2018protecting,darvish2019deepsigns,le2020adversarial,jia2021entangled}. As in classical multimedia watermarking, DNN watermarking includes two stages: \textit{embedding} and \textit{verification}. In the \textit{embedding} stage, the DNN model owner inserts a secret watermark (e.g., signature or a trigger) into the model during the training phase. Existing watermarking techniques can be categorized as either \textit{white-box} or \textit{black-box} based on how much knowledge is available during the \textit{verification} stage. White-box techniques assume the model parameters are available~\cite{uchida2017embedding,darvish2019deepsigns,wang2022integrity}. They insert a string of bits (signature) into the model parameter space via several regularization terms. The ownership of the IP could be claimed when the retrieved string of bits from the suspect model matches the owner's signature. Black-box techniques only have access to model predictions during verification. They leverage backdoor attacks~\cite{gu2019badnets,gao2020backdoor} to embed a watermark (backdoor samples) into the ownership model during the training process, where the class of each backdoor sample is relabelled to a secret class~\cite{le2020adversarial,zhang2018protecting}. The ownership could be verified by querying the suspect model using the predefined backdoor samples and receiving the correct secret class for each sample. 

%The watermarking scheme embeds watermark such as trigger sets into the DNN model so that the model can be verified by extracting the watermark. The watermarking scheme is divided into a method of embedding a specific key into an input value\cite{zhang2018protecting,adi2018turning}, the method of embedding a specific key into a parameter\cite{chen2018deepmarks,darvish2019deepsigns,uchida2017embedding}, or the method of using adversarial images\cite{chen2019blackmarks, jia2021entangled,Merrer2019Adversarial}. Adi et. al. \cite{adi2018turning} proposed watermark as images from other domain and Zhang et. al \cite{zhang2018protecting} generates watermark as images that contain certain content or noise, or as images from other domain. They embed a watermark by fine tuning with the watermark images, and verify a model by testing the watermark images.
\subheading{DNN fingerprinting.} 
DNN fingerprinting mechanisms have been recently introduced as an alternative approach to verify model ownership via two stages called fingerprint extraction and verification. Fingerprinting methods~\cite{IPGuard,Jingjing2020AFA,lukas2019deep,chen2021copy} are all \textit{black-box} techniques. They are \textit{non-invasive}, as opposed to watermarking techniques that are \textit{invasive}. Rather than altering the training process to inject the watermark, fingerprinting directly retrieves a unique property/feature of the owner's model as its fingerprint. The ownership can then be validated if the fingerprint matches with the one extracted from the suspect model. In general, there are two streams of work under this category: \textit{single} and \textit{multiple} fingerprinting. Single fingerprinting uses one feature/property as an identifier. For example, IPGuard~\cite{IPGuard} uses data points close to the model's decision boundaries as that identifier. Lukas \textit{et al.}~\cite{lukas2019deep} propose a conferrable adversarial example that transfers a target label from a source model to its stolen model. They use that as a model identifier. Multiple fingerprinting leverages multiple features/metrics as a fingerprint to handle different types of model stealing and adaptive attacks. For example, Chen et al.~\cite{chen2021copy}  recently introduced \deepjudge, a multi-level metrics mechanism that can be used as a DNN model fingerprinting technique.

%The DNN model has a unique property, fingerprint, for each model. There are papers that want to classify the model and claim the IP of the model using these properties\cite{IPGuard},\cite{Jingjing2020AFA},\cite{sensitive_fingerprint},\cite{deepjudge}. Most papers measured the similarity of the model by generating an input that can distinguish the model well and observing the results of the input value. In the case of deepjudge\cite{deepjudge}, various metrics to quantify the properties of the model are provided to provide a more effective fingerprinting tool.

Existing DNN watermarking and fingerprinting techniques often suffer from two main limitations: architecture dependence and the inability to detect models trained on a combination of stolen and other datasets. In real-world settings, transfer learning and fine-tuning allow stolen models to be retrained on new datasets. This paper empirically demonstrates that \sysname is a robust technique against nine attack scenarios, including those that exploit transfer learning and fine-tuning.

%Although the above streams protect the model IP with high performance, they suffer from two main limitations. \hl{Firstly, they are architecture-dependent by design. For example, training thesame (stolen) dataset on three different DNNs cannot be identified as IP violation, even though all three models absorbed the same dataset ownership IP.} Secondly, due to being architecture-dependent, they struggle to detect transfer learning attacks. For instance, if a pre-trained DNN is stolen and used for transfer learning to a different domain, this cannot be tracked as stolen IP. In other words, they could not track the dataset ownership IP obtained from a dataset across various architectures. Therefore, we propose \sysname, a robust dataset ownership IP tracking technique against 6 attacks.

%\sharif{1 paragraph about DNN fingerprinting as we could be categorised as them.  DeepJudge paper \cite{deepjudge} is the latest in this track but in their intro they cited 2 papers. They said they use single feature which is limited and deepjudge used multiple. We might need to highlight our the difference?}
%-------------------------------------------------------------------------------
\section{Conclusion}
%-------------------------------------------------------------------------------
We propose a novel DNN fingerprinting technique, \sysname, that uses adversarial perturbations in the Fourier frequency domain to effectively identify DNN models. We found that the spectra of gradient-based adversarial perturbations on DNNs can capture unique characteristics of models trained on a specific dataset. To demonstrate the efficacy of our approach, we conducted a comprehensive evaluation of \sysname's detection accuracy on three datasets and three model architectures, subject to various attack scenarios, including multi-model architectures, data augmentation, transfer learning, fine-tuning, pruning, transfer learning with pretrained model, and adversarial learning. Our experimental results show that \sysname is highly robust against these attacks.

%Evaluating a suspect model requires generating and examining adversarial DFT samples with \sysname. Future work could explore ways to generate adversarial samples in a black-box setting.

%Future work
%To evaluate the suspect model, \sysname needs access to the suspect model to be able to generate adversarial DFT samples against that model before examining them with our classifier. A possible future direction could be to explore ways to generate adversarial samples in a black-box setting to provide more flexibility and generalization.

% %-------------------------------------------------------------------------------
\begin{acks}
% %-------------------------------------------------------------------------------
We thank our anonymous shepherd and reviewers for their valuable feedback and insights. Hyoungshick Kim is the corresponding author. This work was supported by Institute for Information \& communication Technology Planning \& Evaluation grant funded by the Korea government (No.2018-0-00532, Development of High-Assurance ($>=$EAL6) Secure Microkernel (50\%), No.2022-0-00995 (30\%), and No.2019-0-01343 (20\%)).
\end{acks}
%\clearpage
\bibliographystyle{ACM-Reference-Format}
\bibliography{reference}

%\clearpage

\appendix
\clearpage
\section{Adversarial Perturbation Images in both the DFT and spatial domains}~\label{Appendix:domain}

This section presents adversarial perturbation images for nine distinct models with three different architectures (ResNet18, VGG16, and DenseNet161) that were trained on three different datasets (CIFAR10, MNIST, and Tiny-ImageNet). These perturbations are presented in both the DFT domain (see Table~\ref{tb:DFT}) and the spatial domain (see Table~\ref{tb:pixel}).

\begin{table}[h]
\centering
\caption{Adversarial perturbation images in the DFT domain for three architectures (ResNet18, VGG16, and DenseNet161) trained on three datasets (CIFAR10, MNIST, and Tiny-ImageNet).}

\begin{tabular}{|c|c|c|c|}
\hline
 & \textbf{ResNet18} & \textbf{VGG16} & \textbf{DenseNet161} \\ \hline
\textbf{CIFAR10} & \makecell{\includegraphics[width=.18\linewidth]{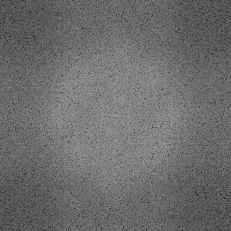}} & \makecell{\includegraphics[width=.18\linewidth]{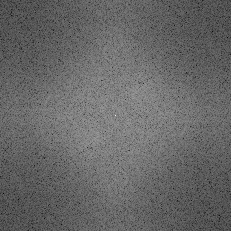}} & \makecell{\includegraphics[width=.18\linewidth]{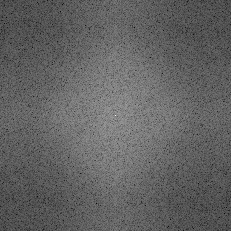}} \\\hline
\textbf{MNIST} & \makecell{\includegraphics[width=.18\linewidth]{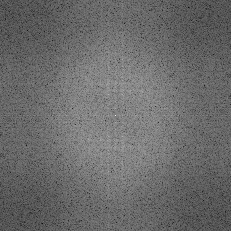}} & \makecell{\includegraphics[width=.18\linewidth]{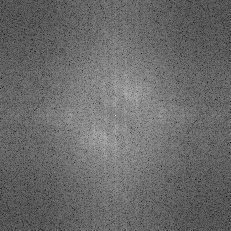}} & \makecell{\includegraphics[width=.18\linewidth]{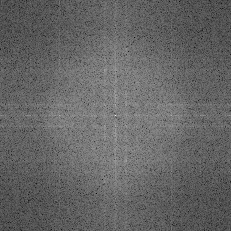}} \\\hline
\begin{tabular}[c]{@{}c@{}} \textbf{Tiny-}\\ \textbf{ImageNet}  \end{tabular} & \makecell{\includegraphics[width=.18\linewidth]{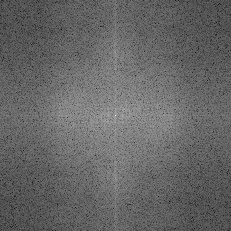}} & \makecell{\includegraphics[width=.18\linewidth]{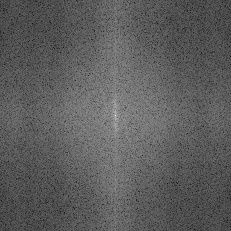}} & \makecell{\includegraphics[width=.18\linewidth]{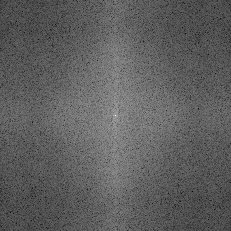}} \\\hline
\end{tabular}
\label{tb:DFT}
\end{table}

\begin{table}[h!]
\centering
\caption{Adversarial perturbation images in the spatial domain for three architectures (ResNet18, VGG16, and DenseNet161) trained on three datasets (CIFAR10, MNIST, and Tiny-ImageNet).}
\begin{tabular}{|c|c|c|c|}
\hline
 & \textbf{ResNet18} & \textbf{VGG16} & \textbf{DenseNet161} \\ \hline
\textbf{CIFAR10} & \makecell{\includegraphics[width=.18\linewidth]{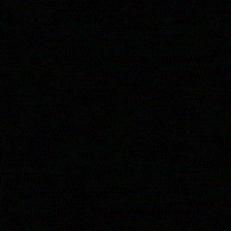}} & \makecell{\includegraphics[width=.18\linewidth]{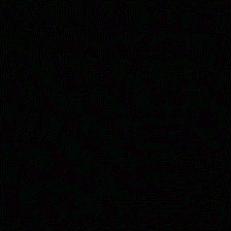}} & \makecell{\includegraphics[width=.18\linewidth]{figure/dft1/cifar10_densenet.jpg}} \\\hline
\textbf{MNIST} & \makecell{\includegraphics[width=.18\linewidth]{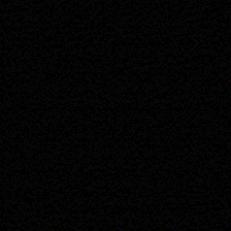}} & \makecell{\includegraphics[width=.18\linewidth]{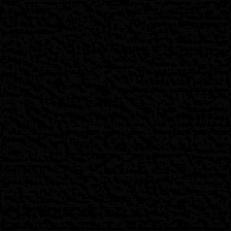}} & \makecell{\includegraphics[width=.18\linewidth]{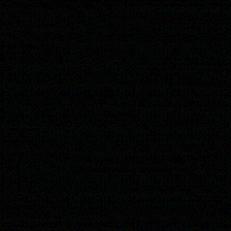}} \\\hline
\begin{tabular}[c]{@{}c@{}} \textbf{Tiny-}\\ \textbf{ImageNet}  \end{tabular} & \makecell{\includegraphics[width=.18\linewidth]{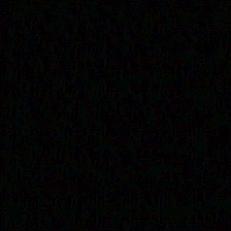}} & \makecell{\includegraphics[width=.18\linewidth]{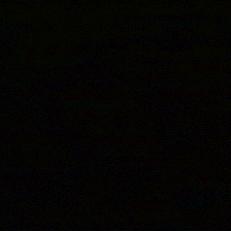}} & \makecell{\includegraphics[width=.18\linewidth]{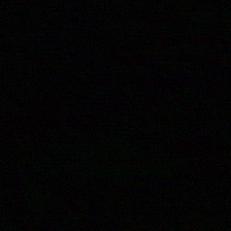}} \\\hline
\end{tabular}
\label{tb:pixel}
\end{table}

We evaluated the similarity between adversarial images within the same dataset but different architectures using the Mean Squared Error (MSE) metric in both the DFT and spatial domains. Specifically, we analyzed 18 different adversarial images generated from nine models, consisting of all combinations of three distinct architectures (ResNet, VGG, and DenseNet) and three different datasets (CIFAR10, MNIST, and Tiny-ImageNet), with nine images for each domain. Our results showed that the average MSE of images from models trained on the same dataset was 0.0074 in the DFT domain, which is about seven times lower than the average MSE of 0.0505 found in the spatial domain. 

\section{Datasets used in experiments}\label{App:experiment_setup}

In Section~\ref{sec:Experiments}, we conduct experiments on the four image classification datasets. Table~\ref{tb:dataset} described the number of classes and usage of each dataset. 

\begin{table}[h!]
\centering
\caption{Description of the datasets for experiments.}
\resizebox{2.0in}{!}{
\begin{tabular}{lll}
\hline
Dataset & \# Classes & Usage \\ \hline
CIFAR10 & 10& Victim / Suspect\\ %\hline
MNIST & 10 & Victim / Suspect\\ %\hline
Tiny-ImageNet & 100 & Victim / Suspect\\ %\hline
ImageNet& 1000& Suspect \\ \hline
\end{tabular}
}
\label{tb:dataset}
\end{table}

\section{Models used in experiments}\label{App:experiment_model}

Table~\ref{tb:model_details} reports the number of parameters and accuracy of models used in the experiment in Section~\ref{sec:Experiments}.

\begin{table}[h!]
\centering
\caption{Datasets, models, and parameters we used and \new{mean values along with the standard deviation values of baseline accuracy.}}
\resizebox{2.6in}{!}{
\begin{tabular}{llll}
\hline
Dataset & Architecture & \# Params & Accuracy (\%) \\ \hline
& ResNet18&  11181642& \new{74.15 (0.37)}\\ \cline{2-4} 
CIFAR10 & VGG16 &134301514& \new{82.62 (2.81)}\\ \cline{2-4} 
& DenseNet161& 26494090& \new{70.80 (0.82)}\\ \hline
& ResNet18 & 11181642& \new{99.47 (0.04)}\\ \cline{2-4}
MNIST &  VGG16& 134301514 & \new{99.47 (0.04)}\\ \cline{2-4} 
& DenseNet161& 26494090&  \new{99.26 (0.08)}\\ \hline
& ResNet18 & 11181642&  \new{35.27 (0.63)}\\ \cline{2-4}
Tiny-ImageNet &  VGG16& 134301514&  \new{39.46 (0.40)}\\ \cline{2-4} 
& DenseNet161& 26494090&  \new{33.13 (2.18)}\\ \hline
\end{tabular}}
\label{tb:model_details}
\end{table}
%-------------------------------------------------------------------------------
%\clearpage

\section{Effects of training dataset size and dimensions}
\label{Appendix:Effects of training dataset size and dimensions}
%\sharif{Explain why we select 9600 samples ...}

The performance of a classifier may depend on the training dataset. Generally, a larger training dataset might facilitate the production of a higher-performance model. In our case, generating large adversarial DFTs samples dataset might mean higher time cost. For generating a balanced model, we test the relationship between performance and dataset size. The training dataset is generated with various sizes: 2400, 4800, 7200, and 9600 images. These four training datasets are generated from ResNet18, VGG16, and DenseNet161 models trained on CIFAR10, and we use the same seed dataset of adversarial DFT samples for consistency.
The balanced accuracy results of the classifier are 97.28\%, 98.50\%, 96.28\%, and 96.82\%, respectively, when the training dataset size is 2400, 4800, 7200, and 9600. Based on these results, we recommend 4800 for the training dataset size for \sysname. 

We also observe that the performance of \sysname can vary depending on the adversarial image dimensions. The smaller the size of the adversarial image, the smaller the perturbation that could be captured from the model dataset intelligence, and the lower the performance of the \sysname might be. If the size of the image is $32\cdot32\cdot3$, the model exhibits almost indistinguishable performance, but if the size of the image is $224\cdot224\cdot3$, as currently used in the experimental setting, the detection performance is high. Therefore, we recommend generating large-dimensional adversarial images when using \sysname.

%-------------------------------------------------------------------------------
%\clearpage

\section{Effects of threshold}~\label{Appendix:thresholding}

\begin{figure}[h!]
    \centering
    \begin{tabular}{cc}
        \includegraphics[width=.77\linewidth]{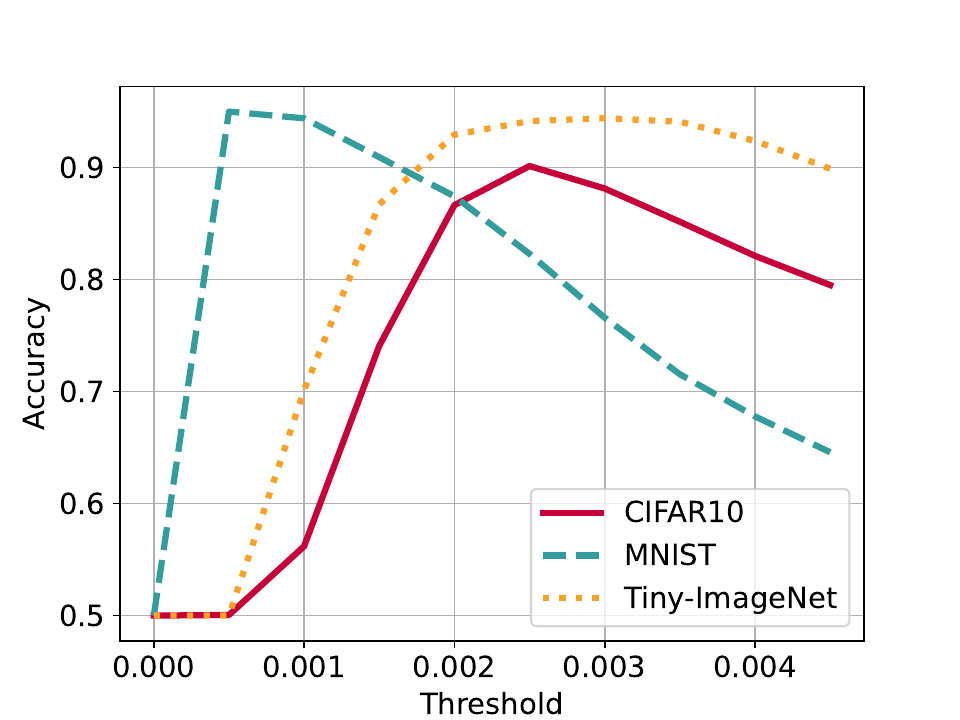}
    \end{tabular}
    \caption{Performance of classifiers with thresholds.}
    \label{fig:thre}
\end{figure}

\sysname is sensitive to how the threshold is determined. Figure~\ref{fig:thre} shows the balanced accuracy of the classifier for each dataset with their corresponding thresholds. 

\section{Experimental results of \sysname with a large-sized model architecture}\label{app:large}

We evaluated the effectiveness of \sysname with a large-scale model architecture (AlexNet) by building a classifier using the Tiny-ImageNet as the victim dataset. As shown in Table~\ref{tb:Alexnet}, the classifier can effectively detect all 12 suspect models correctly.

When using AlexNet as the architecture, we observed a high theft image rate of over 95\% for both stolen and benign cases. This suggests that \sysname operates effectively with large-scale model architectures such as AlexNet.

\begin{table}[b]
\centering
\caption{Performance of \sysname with AlexNet when the victim dataset is Tiny-ImageNet.}
\resizebox{3.2in}{!}{
\begin{tabular}{c|c|c|c|c}
\hline
Suspect & \begin{tabular}[c]{@{}c@{}} ground\\truth \end{tabular}& architecture & \begin{tabular}[c]{@{}c@{}} Theft Image\\ Rate \end{tabular} & \begin{tabular}[c]{@{}c@{}} Copy \\Detection(\%) \end{tabular}\\ \hline
\multirow{4}{*}{\begin{tabular}[c]{@{}c@{}} Tiny-\\ImageNet  \end{tabular}} & \multirow{4}{*}{stolen} & Alexnet &  95.83& \new{100}  \\ \cline{3-4}
 &  & ResNet18 & 94.10 & \new{100} \\ \cline{3-4}
 &  & VGG16 &  93.06 & \new{100}  \\ \cline{3-4}
 &  & DenseNet161 & 94.79 &   \new{100}  \\ \cline{1-5}
\multirow{4}{*}{\begin{tabular}[c]{@{}c@{}} CIFAR10 \end{tabular}} & \multirow{4}{*}{benign} & Alexnet &  100 &  \new{100}  \\ \cline{3-4}
 &  & ResNet18 & 75.69 & \new{100}  \\ \cline{3-4}
 &  & VGG16 &71.53 &  \new{100}  \\ \cline{3-4}
 &  & DenseNet161 & 74.31 &   \new{100}  \\ \cline{1-5}
\multirow{4}{*}{\begin{tabular}[c]{@{}c@{}} MNIST\end{tabular}} & \multirow{4}{*}{benign} & Alexnet &100 & \new{100}  \\ \cline{3-4}
 &  & ResNet18 & 100 &   \new{100}   \\ \cline{3-4}
 &  & VGG16 & 100 &   \new{100}  \\ \cline{3-4}
 &  & DenseNet161 & 76.39 &  \new{100}   \\ \hline
\end{tabular}}
\label{tb:Alexnet}
\end{table}

\end{document}